\documentclass[%
 reprint,
superscriptaddress,
 amsmath,amssymb,
 aps,
 pre,
]{revtex4-2}

\usepackage{graphicx}
\usepackage{bm}

\usepackage{algorithm}
\usepackage{algpseudocode}
\usepackage{xcolor}

\begin{document}


\title{Effect of Constraint Relaxation on the Minimum Vertex Cover Problem in Random Graphs}

\author{Aki Dote}
\affiliation{
 Graduate School of Arts and Sciences,
 The University of Tokyo,
 Komaba, Meguro-ku, Tokyo 153-8902, Japan
}
\affiliation{
 Fujitsu Limited.
 4-1-1 Kamikodanaka, Nakahara-ku, Kawasaki, 211-8588, Japan
}

\author{Koji Hukushima}
\affiliation{
 Graduate School of Arts and Sciences,
 The University of Tokyo,
 Komaba, Meguro-ku, Tokyo 153-8902, Japan
}
\affiliation{
Komaba Institute for Science, The
University of Tokyo, 3-8-1 Komaba, Meguro-ku, Tokyo 153-8902, Japan
}

\date{\today}

\begin{abstract}
A statistical-mechanical study of the effect of constraint relaxation on the minimum vertex cover problem in Erd\H{o}s-R\'enyi random graphs is presented.
Using a penalty-method formulation for constraint relaxation, typical properties of solutions, including infeasible solutions that violate the constraints, are analyzed by means of the replica method and cavity method. 
The problem involves a competition between reducing the number of vertices to be covered and satisfying the edge constraints. 
The analysis under the replica-symmetric (RS) ansatz clarifies that the competition leads to degeneracies in the vertex and edge states, which determine the quantitative properties of the system, such as the cover and penalty ratios.
A precise analysis of these effects improves the accuracy of RS approximation for the minimum cover ratio in the replica symmetry breaking (RSB) region.
Furthermore, the analysis based on the RS cavity method indicates that the RS/RSB boundary of the ground states with respect to the mean degree of the graphs is expanded, and the critical temperature is lowered by constraint relaxation. 
\end{abstract}

\maketitle


\section{Introduction} \label{sec:level1}
Combinatorial optimization problems play an important role in academic fields such as mathematics, physics, and chemistry, in that several problems in each field can be reduced to optimization problems. 
In industrial fields such as manufacturing, logistics, and finance, real-world problems are often formulated as optimization problems.
There is an increasing demand for efficient tools to solve these problems.
On an algorithmic view, several physics-inspired optimization algorithms, such as simulated annealing~\cite{kirkpatrick1983optimization} and quantum annealing~\cite{kadowaki1998quantum,farhi2001quantum}, have emerged. 
Simultaneously, from a hardware perspective, triggered by the development of the quantum annealer~\cite{johnson2011quantum}, several Ising machines based on different physical phenomena have been developed in the last decade~\cite{yamaoka201520k,matsubara2020digital,mcmahon2016fully,goto2019combinatorial,mohseni2022ising}.
These developments have significantly expanded the methodologies and scope of combinatorial optimization.

The constrained combinatorial optimization problems discussed in this study aim at finding a combination of discrete values that minimizes or maximizes a cost function among feasible solutions, that is, a set of values satisfying given constraints.
Typical problems include the traveling salesperson problem (TSP), minimum vertex cover (MVC) problem, and knapsack problem.
Certain types of problems, such as the TSP, can allow efficient exploration of optimal or near-optimal solutions among the feasible solutions by appropriately considering their structured constraints. 
However, it is challenging to systematically generate feasible solutions to problems with arbitrary constraints.
Particularly, in industrial applications, where problems often combine multiple constraints, it is difficult to search only for feasible solutions depending on the structure of each problem instance.
The difficulties associated with constraints become more pronounced when utilizing dedicated hardware such as quantum and classical Ising machines. 
Although these machines are powerful, they are limited in their ability to directly address intricate constraints, except for some specific types~\cite{lucas2014ising,mohseni2022ising}.

A well-established method to address such constrained combinatorial optimization problems is the penalty method~\cite{smith1997penalty, coello2002theoretical}. 
This method constructs an energy function by adding non-negative constraint functions as a penalty to the cost function to be minimized. 
Then, feasible solutions are obtained by finding the lowest energy state of this energy function. 
To ensure that the lowest energy state satisfies the constraints, the penalty coefficients of the constraints should be sufficiently large. 
It is known that smaller penalty coefficients can practically lead to faster solutions for combinatorial optimization problems~\cite{smith1997penalty}. 
However, determining an appropriate penalty coefficient in advance for any given problem instance remains a challenging task.  
Therefore, adaptive penalty functions~\cite{glover1993user} and other heuristic methods have been introduced to dynamically adjust penalty coefficients. 
These adjustments are highly problem-dependent, and there is a lack of theoretical analyses.

The theoretical analysis of combinatorial optimization problems using statistical mechanics originates from the application of methods of statistical physics for random systems to the computational complexity theory~\cite{fu1986application}. 
Particularly, statistical-mechanical ideas have proven invaluable for analyzing the typical-case behavior of the randomized problems~\cite{mezard1986replica,fontanari1995statistical}.
In this context, intriguing insights have emerged from the analysis of the boundary between problem instances that satisfy or violate constraints~\cite{kirkpatrick1994critical,monasson1997statistical}. 
These boundaries often highlight the transition in computational complexity changes from polynomial time to exponential time~\cite{weigt2000number,selman1996critical}, and the region where approximate algorithms yield exact optimal solutions~\cite{bauer2001core,mezard2003two}.
These results are closely related to phase-transition phenomena and provide a deeper understanding of the structure of the problems~\cite{kirkpatrick1994critical,monasson1997statistical,weigt2000number,selman1996critical,bauer2001core,mezard2003two,weigt2001minimal,monasson19992+,zhou2003vertex,zhang2009stability,zdeborova2009statistical}.
It should be emphasized that, with a few exceptions~\cite{mezard1986replica}, previous statistical-mechanical analyses have focused either on combinatorial optimization problems restricted to feasible solutions or on constraint-satisfaction problems without cost functions. 
Constrained optimization problems, particularly those with relaxed constraints such as the penalty method, have not been extensively analyzed.

The purpose of this study is to explore the implications of constant relaxation in the penalty method by using statistical mechanics analysis. 
We focus on the MVC problem, extensively studied in statistical mechanics and recognized as one of the most fundamental \textsf{NP}-hard problems~\cite{karp1972reducibility}. 
Specifically, our work extends previous studies of the MVC problems on Erd\H{o}s-R\'enyi (ER) random graphs~\cite{erdHos1960evolution}, analyzing ground states with replica methods~\cite{weigt2001minimal} and finite-temperature properties using the cavity method~\cite{zhang2009stability}.
It should be noted that the MVC problem is equivalent to the maximum independent set problem and the maximum clique problem, which have broad practical applications. 
These problems have also been studied from a statistical mechanics perspective in the context of random graphs~\cite{dall2009statistical,ding2016maximum}. 
In contrast to the previous studies, our approach considers systems with infeasible solutions that do not satisfy all constraints under finite constraint strength and includes previous studies as a limit with infinite constraint strength. 
By analyzing replica-symmetric (RS) solutions, we obtain conditions for obtaining feasible solutions in the low-temperature limit.
This analysis also reveals the competing structures of the cost function and constraint strength in the lowest-energy states under infeasible conditions.
Furthermore, through a stability analysis of the RS solution and cavity-method analysis, we determine the transition temperature $T_{\mathrm{c}}$ of the replica symmetry breaking (RSB), and this temperature decreases as the constraint strength is decreased. 
Finally, numerical experiments demonstrate indications of RSB at finite temperatures and the lack of self-averaging properties.

The paper is organized as follows.
In Sec.~\ref{sec:models}, we define the statistical mechanical model of MVC with the penalty method and conduct a finite temperatures analysis using the replica method.
Next, Sec.~\ref{sec:ground_state} provides a detailed analysis of the ground-state properties of the system, particularly, the infeasible solutions that arise when the constraints are relaxed.
In Sec.~\ref{sec:stabilityofselfconsistent}, we analyze the stability of the self-consistent equations for the RS solutions.
In Sec.~\ref{sec:stabilitybycavity}, we further explore the stability of the RS solutions at finite temperatures, particularly their constraint dependence, by using the cavity method.
In Sec.~\ref{sec:numericalresults}, the stability analysis described in the previous section is verified, and the behavior of the system, which is not fully understood by the analysis of the RS solution, is investigated by Markov-chain Monte Carlo (MCMC) simulations.
Finally, conclusions are presented in Sec.~\ref{sec:conclusion}. 
In the appendices, we present derivations of some mathematical formulations, population dynamics methods, and their behavior at low temperatures.

\section{\label{sec:models}Model and statistical mechanical formulation}
This section provides an overview of the statistical-mechanics analysis based on the replica method for the MVC problem, as discussed in Refs.~\cite{weigt2000number,weigt2001minimal}. 
We explore its extension to problems with relaxed constraints and discuss the resulting properties. 

\subsection{MVC problem}
\label{Sec:MVCandPen}
Given an undirected graph $G(V, E)$ with $N$ vertices $V$ and edges $E$, MVC describes the problem of finding the minimum subset of the vertices $V_c\subset{V}$ that covers the graph $G$, where ``cover" means that at least one of the two vertices connected by each edge belongs to $V_c$. 

Let $x_i=1$ with $i\in\{1,2,\dots,N\}$ denote that the $i$-th vertex is covered and $x_i=0$ denote that the $i$-th vertex is uncovered. 
The adjacency matrix of a graph $G$ is denoted by $\bm{c}$, and the matrix element $c_{ij}$ is given by $1$ if the edge $(ij)$ is connected and $0$ otherwise. 
This problem can be formulated as a constrained combinatorial optimization problem for $\bm{x}=(x_1,\cdots,x_N)$ with
\begin{equation}
    \begin{aligned}
        &\text{minimize: } M(\bm{x}) = \sum_{i=1}^N x_i ,\\
        &\text{s.t.: } V(\bm{x}; G) = \sum_{(ij)}c_{ij}(1-x_i)(1-x_j) = 0, 
    \end{aligned}
    \label{eqn:problem}
\end{equation}
where $M(\bm{x})$ and $V(\bm{x}; G)$ are called the cost function and the penalty function, respectively.
The penalty function $V(\bm{x}; G)$ represents the constraints for a given graph $G$. 
An assignment $\bm{x}$ that satisfies the constraint conditions is called a feasible solution, and one that does not satisfy is called an infeasible solution.
With this formulation, the infeasible solution is characterized by  $V(\bm{x}; G) > 0$.

Generally, there are two main approaches to solving constrained combinatorial optimization problems. 
One is to restrict the search for the optimal solution to feasible solutions only.
This approach is efficient when feasible solutions can be generated systematically without missing any feasible solutions. However, this approach is applicable only to a specific class of established problems.
The other is to search for the optimal solutions, including infeasible solutions.
This is less efficient for problems to which the former method can be applied, but can be applied to any problem for which it is difficult to find feasible solutions.

To analyze the typical properties of MVC problems, including infeasible solutions, we focus on a formulation based on the penalty method, which is one of the simplest methods of the latter. 
The energy function incorporating the constraints into the penalty function is defined as
\begin{equation}
    E(\bm{x}; G) = \mu M(\bm{x}) + \gamma V(\bm{x}; G) ,
    \label{eq:energywithpenalty}
\end{equation}
where $\mu$ is a parameter with dimensions of energy, and $\gamma$ is a positive penalty coefficient that determines the strength of the constraint.
Without loss of generality, $\mu$ is used as the unit of the energy hereafter with $\mu=1$.
For sufficiently large $\gamma$, the solution $\bm{x}$ that minimizes $E(\bm{x}; G)$ satisfies the constraint and is then the optimum solution to the problem of Eq.~(\ref{eqn:problem}). 
The search process in the limit of $\gamma\to\infty$ realizes only feasible solutions that completely satisfy $V(\bm{x}; G)=0$.
It has been suggested that the value of $\gamma$ should be as small as possible in order to obtain a solution in a short time~\cite{smith1997penalty}. 
However, it is generally difficult to determine the appropriate value of $\gamma$ in advance.
Therefore, heuristic methods of dynamically changing $\gamma$ and methods of finding the optimal $\gamma$ value have been proposed to find a solution efficiently~\cite{coello2002theoretical}.
One example is the adaptive penalty function, which dynamically controls $\gamma$ and efficiently searches for feasible solutions by moving the state back and forth between feasible and infeasible regions.
As a concrete example, an optimization method using the tabu search with multiple penalty functions is described in Ref.~\cite{glover1993user}.

In the case of MVC, the condition for the minimum-energy solutions to satisfy the constraints is $\gamma>1$~\cite{lucas2014ising}.
This is intuitively obvious: when $\gamma=1$, reducing the cover subset by one and violating the constraint by one make exactly the same energy contribution.
Thus, by setting the value of $\gamma>1$, the minimum-energy solution satisfies the constraint and gives a minimum cover subset.
Note that the energy for $\gamma=1$ is equal to the minimum coverage.
If $\gamma$ is dynamically changed during the search process, it should eventually return to $\gamma > 1$, and we discuss the possibility of utilizing $\gamma < 1$ in the path during the search. 
For this purpose, it is necessary to clarify the nature of the finite $\gamma$ region, including $\gamma=1$, which extends the previous statistical-mechanical study of MVC at ``finite temperatures" with $\gamma=\infty$~\cite{weigt2001minimal}.

\subsection{Statistical mechanics of combinatorial optimization problems with constraints}

When analyzing combinatorial optimization problems using statistical mechanics, we introduce a probability distribution that follows the solution $\bm{x}$, referred to as the ``state" in the physical terminology below. 
First, the function to be minimized is considered to be the energy, and the equilibrium distribution of the state $\bm{x}$ at the inverse temperature $\beta$ is assumed to be the canonical distribution. 
In the context of statistical mechanics, the low-temperature limit, $\beta\to\infty$, is generally considered to obtain the optimal solution as the ground state of the system. 

Using the energy function of Eq.~(\ref{eq:energywithpenalty}), the partition function of MVC for given $G$ is defined as
\begin{equation}
    Z(\beta,\gamma;G) = \sum_{\bm{x}} e^{-\beta (\mu M(\bm{x}) + \gamma V(\bm{x};G))}, 
    \label{eq:partitionfunction}
\end{equation}
where the sum is taken for all states of $\bm{x}\in\{0,1\}^N$, including feasible and infeasible states. 
In previous studies~\cite{mezard1986replica, fontanari1995statistical,monasson1997statistical,weigt2001minimal,zdeborova2006number,zhou2003vertex,zhang2009stability} the state $\bm{x}$ was restricted to only feasible states.
Assuming $\mu M(\bm{x})$ as the energy function, the partition function is given by 
\begin{equation}
    Z(\beta;G) = \sum_{\bm{x}} e^{-\beta\mu M(\bm{x})} \delta(V(\bm{x};G),0) ,
    \label{eq:partionfunction_feasible}
\end{equation}
where $\delta$ is the Kronecker delta function that restricts the sum considering only feasible states.
By taking the limit $\gamma\to\infty$ with $\beta>0$, the partition function of Eq.~(\ref{eq:partitionfunction}) is reduced to that of Eq.~(\ref{eq:partionfunction_feasible}). 
It should be noted that for $\beta=0$, the sum in Eq.~(\ref{eq:partitionfunction}) takes all states with equal weight, 
whereas that in Eq.~(\ref{eq:partionfunction_feasible}) takes only the feasible states. 
Thus, the entropy in the high-temperature limit is different for the two systems.

As described above, we discuss here the system defined by Eq.~(\ref{eq:partitionfunction}). 
Then, the free energy density $f$ is expressed as 
\begin{equation}
  -\beta f(\beta,\gamma;G) = \frac{1}{N}\ln Z(\beta,\gamma; G).
\end{equation}
Using the free energy density, thermal averages at inverse temperature $\beta$ of the energy density $\varepsilon$, the cover ratio $\rho$, which is the density of $x_i = 1$, and penalty ratio $\nu$ are obtained respectively as 
\begin{align}
  \varepsilon(\beta,\gamma;G) = \left\langle\frac{E}{N}\right\rangle = \frac{\partial \beta f}{\partial \beta}, \label{eq:energy_density}\\
  \rho(\beta,\gamma;G) = \left\langle\frac{M}{N}\right\rangle = \frac{\partial f}{\partial \mu}, \label{eq:cost_density}\\
  \nu(\beta,\gamma;G) = \left\langle\frac{V}{N}\right\rangle = \frac{\partial f}{\partial \gamma} ,
  \label{eq:penalty_density}
\end{align}
where $\langle\cdots\rangle$ denotes the thermal average.
From Eq.~(\ref{eq:energywithpenalty}), it is obvious that $\varepsilon=\rho+\gamma\nu$. 
Note that $\nu$ is also a non-increasing function of $\gamma$, just as $\varepsilon$ is a non-increasing function of $\beta$. 

Taking $\beta\to\infty$ yields the minimum energy density, and when $\nu=0$, the minimum cover ratio is obtained as $e=\rho$, which is the solution of MVC. 
As discussed below in this section, when $\gamma\le 1$, since $\nu$ is positive even at $\beta\to\infty$, $\varepsilon$ is the minimum energy, but $\rho$ is not the minimum cover ratio.
One of the main purposes of this study is to analyze the behavior of the system for finite $\gamma$. 

\subsection{Random graph and replica trick}

It is challenging to calculate the partition function or free energy for any given graph instance $G$ for large $N$.
If the typical properties of ensembles of instances of this problem, rather than the individual graph instances, are to be determined, the statistical mechanics of random systems can be used by introducing a random-graph ensemble controlled by a few parameters. 
Here, we consider the ER random graph~\cite{erdHos1960evolution} as an ensemble of random instances in which the matrix elements of the adjacency matrix are given by the probability distribution defined as
\begin{equation}
 P(\bm{c}) = \prod_{(ij)}\left(\frac{c}{N}\delta(c_{ij},1)+\left(1-\frac{c}{N}\right)\delta(c_{ij},0)\right).
 \label{eq:ER}
\end{equation}
This random graph is a sparse graph with mean degree $c$, and when $N$ is sufficiently large, the distribution of degrees follows the Poisson distribution, $P(k)=e^{-c}c^k/k!$.

The average for the graph ensemble of the free-energy density in the thermodynamic limit is expressed as 
\begin{equation}
  [f]_G = -\lim_{N\rightarrow\infty }\frac{1}{\beta N}\left[\ln Z(\beta,\gamma; G)\right]_G .
  \label{eq:thermodynamic_limit}
\end{equation}
where $[\cdots]_G$ denotes the graph average for $P(\bm{c})$. 
From the averaged free-energy density, typical expected values of the cover and penalty ratios can be calculated.
To take the graph average, we use the well-known replica trick~\cite{mezard1987spin} represented by 
\begin{equation}
    \left[\ln Z\right]_G
    = \lim_{n\rightarrow 0}\frac{1}{n}\ln\left[Z^n\right]_G.
    \label{eq:replica_trick}
\end{equation}
This is a mathematically correct identity at the limit of the real number $n$.
Assuming $n$ to be an integer, the replica trick first calculates the partition function of $n$-replicated system, which is relatively easy to compute, and then takes the limit of $n\to0$ through analytic continuation.

The replicated partition function of MVC for $n$ replicas with graph averaging reads
\begin{align}
 [Z^n]_G =& \sum_{\{\bm{x}^{(\alpha)}\}}e^{-\beta\mu\sum_{\alpha}M(\bm{x}^{(\alpha)})}
 \left[e^{-\beta\gamma \sum_{\alpha} V(\bm{x}^{(\alpha)})}\right]_G \nonumber\\
 =& \sum_{\{\vec{x}_i\}}e^{-\beta\mu\sum_{i}\vec{1}\cdot\vec{x}_i} e^{-\beta\gamma \sum_{i,j} \mathcal{V}(\vec{x}_i,\vec{x}_j)},
 \label{eq:eff_Z}
\end{align}
where the sum of the replica index $\alpha$ is taken from $1$ to $n$, $\vec{x}_i=(x_i^{(1)},\dots,x_i^{(n)})$ is a replica vector for the $i$-th vertex, 
and an interaction term between replicas is given by 
\begin{equation}
    \mathcal{V}(\vec{x}_i,\vec{x}_j)
    = \frac{c}{2N}\left(1-e^{-\beta\gamma(\vec{1}-\vec{x}_i)\cdot(\vec{1}-\vec{x}_j)}\right). 
\end{equation}
To proceed with the calculation, we employ the order parameter introduced by Monasson~\cite{monasson1997statistical}, defined as 
\begin{equation}
 C(\vec{\xi}) = \frac{1}{N}\sum_i\prod_{\alpha=1}^n\delta\left(\xi^{(\alpha)},x_i^{(\alpha)}\right) .
\end{equation}
This is a distribution function with $2^n$ elements that represents the fraction of $N$ replica vectors $\{\vec{x}_i\}$ that coincide with $\vec{\xi}$, satisfying the normalization condition $\sum_{\vec{\xi}}C(\vec{\xi})=1$.

By summing over $\{\vec{x}_i\}$ in Eq.~(\ref{eq:eff_Z}), the averaged free-energy density $[f]_G$ is obtained as 
\begin{equation}
    -\beta[f]_G = \lim_{n\rightarrow 0} \frac{1}{n} \max_{C(\vec{\xi}); \sum_{\vec{\xi}}C(\vec{\xi})=1}
    g\left(\{C(\vec{\xi})\}\right), 
\end{equation}
where $g$, sometimes called free-entropy density~\cite{mezard2009information}, is given by
\begin{widetext}
\begin{equation}
  g(\{C(\vec{\xi})\}) = 
  -\sum_{\vec{\xi}}C(\vec{\xi})\ln C(\vec{\xi})-\beta\mu\sum_{\vec{\xi}}C(\vec{\xi})\vec{1}\cdot\vec{\xi}-\frac{c}{2}\sum_{\vec{\xi},\vec{\xi'}}C(\vec{\xi})C(\vec{\xi'})\left(1-\exp\left(-\beta\gamma(\vec{1}-\vec{\xi})\cdot(\vec{1}-\vec{\xi'})\right)\right).
  \label{eq:freeentropyfunction}
\end{equation}
\end{widetext}
Taking the limit $\gamma\rightarrow\infty$, this expression is reduced to the free-energy density of the previous study~\cite{weigt2001minimal}. 
Generally, the extremum condition for $\{C(\vec{\xi})\}$ is given by $2^n$ saddle-point equations, $\partial g(\{C(\vec{\xi})\})/\partial C(\vec{\xi}) = 0$ for $\forall \vec{\xi}\in\{0,1\}^n$.

\subsection{RS solution}

In this study, we assume that under the RS ansatz, the solution of the saddle-point equations is invariant with respect to the permutation of the replica index. 
The order parameter is replaced by the $n+1$ RS order parameter $C_{\rm RS}(y=\sum_\alpha\xi^{\alpha})$, which can be written in terms of the one-body distribution function $p(h)$ as
\begin{equation}
  C_{\rm{RS}}\left(\sum_\alpha \xi^\alpha\right) = \int dh p(h)\frac{\exp(-\beta h\sum_\alpha \xi^\alpha)}{(1+e^{-\beta h})^n}.
  \label{eqn:eff_potential}
\end{equation}
From the normalization condition of $C_{\rm RS}$, $p(h)$ is also normalized and can be regarded as a probability distribution function of an effective field $h$ acting on $n$ independent bits $\xi^{\alpha}$ at inverse temperature $\beta\mu$.

Substituting Eq.~(\ref{eqn:eff_potential}) into the saddle point equations and taking the replica limit $n\to0$, we obtain the self-consistent equation for $p(h)$ as
\begin{equation}
 p(h) = 
 e^{-c} \sum_{l=0}^\infty \frac{c^l}{l!}\!\!\int \prod_{i=1}^l dh_ip(h_i) \delta\!\!\left(h-1+\sum_{j=1}^l K(h_j;\beta,\gamma)\right). 
\label{eq:self-consistent}
\end{equation}
where the dependence of $\beta$ and $\gamma$ appears explicitly only in $K(h;\beta,\gamma)$ given by 
\begin{equation}
 K(h;\beta,\gamma)=\frac{1}{\beta}\log\frac{1+e^{-\beta h}}{e^{-\beta\gamma}+e^{-\beta h}}.
 \label{eq:Kofh}
\end{equation}
See Appendix~\ref{sec:self_consistent_eq} for details.
This expression includes the result of the previous study~\cite{weigt2001minimal} in the limit of $\gamma\rightarrow\infty$, which corresponds to the case of feasible states only. 
Note that since $K(h;\beta,\gamma)\ge0$ for positive $\beta$ and $\gamma$, the support of $p(h)$ is restricted to $h\le1$, that is, $p(h)=0$ for $h>1$.
The right-hand side of the self-consistent equation contains a delta function, and particularly, the $l=0$ term contributes directly as a sum. 
Therefore, $p(h)$ is not an analytical function, which makes using general functional expansion methods challenging. 
To find numerical solutions to the self-consistent equation, the method of population dynamics~\cite{mezard2001bethe} is generally used, in which $p(h)$ is approximated by its sample set $\{h_i\}$ and solved iteratively. 
See Appendix~\ref{sec:population_dynamics} for details.

The physical quantities averaged with respect to the graph ensemble are given using $p(h)$ under the RS ansatz. 
The cover and penalty ratios are derived directly from the saddle point of Eq.~(\ref{eq:freeentropyfunction}).
See Appendix~\ref{sec:cover_and_penalty_ratio} for details.
As a result, the cover ratio is given by
\begin{equation}
 \rho(\beta,\gamma) = \int dhp(h) \frac{e^{-\beta h}}{1+e^{-\beta h}}, 
 \label{eq:coverratio}
\end{equation}
where the factor in the integral represents the probability that a vertex is covered under the effective field $h$. 
This expression is identical to the result in the previous study~\cite{weigt2001minimal}, but $p(h)$ depends on $\gamma$, not only on $\beta$. 
Similarly, the penalty ratio is given by 
\begin{align}
    \nu(\beta,\gamma) =& \frac{c}{2} \int dhdh' p(h)p(h') \nonumber \\
     \times &\frac{e^{-\beta\gamma}}{e^{-\beta\gamma} + e^{-\beta h} + e^{-\beta h'} + e^{-\beta (h+h')}} , 
     \label{eq:penaltyratio}
\end{align}
where the factor in the integral is interpreted as the probability that neither of two vertices, $\xi$ and $\xi'$, are covered when their effective energies are represented by $h\xi + h'\xi' + \gamma(1-\xi)(1-\xi')$.
The energy density is obtained by the sum of these terms as $\varepsilon(\beta,\gamma) = \rho(\beta,\gamma) + \gamma \nu(\beta,\gamma)$.
We also see that the penalty ratio disappears at $\gamma\to\infty$. 

\begin{figure}[t]
    \includegraphics[width=0.9\linewidth]{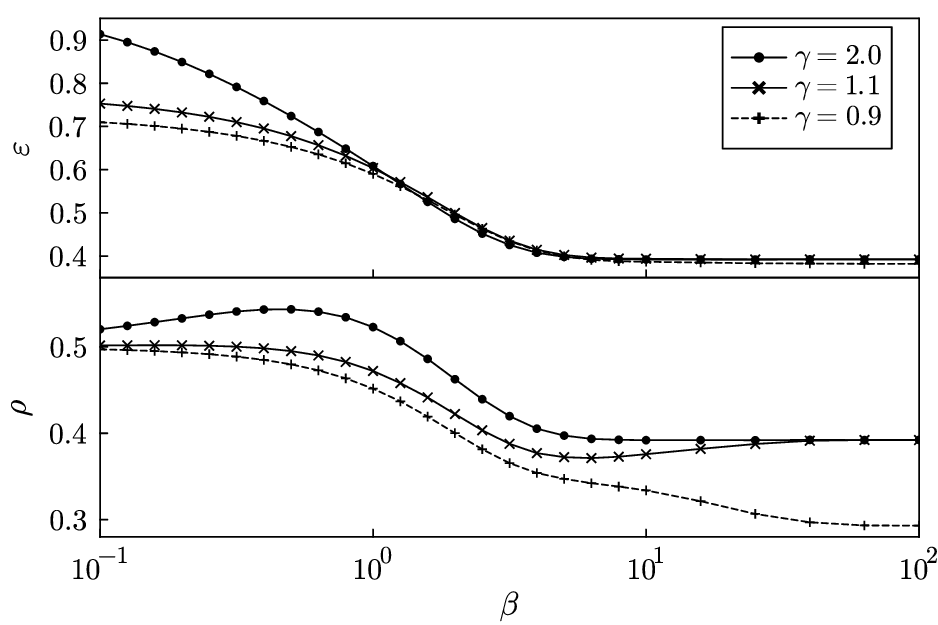}
    \caption{\label{fig:energy_beta_gamma}Inverse-temperature $\beta$ dependence of the energy density $\varepsilon$ and the cover ratio $\rho$  for $\gamma=2.0$, $1.1$ and $0.9$ for $c=2.0$. For $\gamma>1$ (solid lines), both $\varepsilon$ and $\rho$ are asymptotically converged to the minimum cover ratio for large $\beta$, but for $\gamma<1$ (dashed), $\varepsilon$ and $\rho$ are converged to lower values. These calculations were obtained by the method of population dynamics with $10^5$ populations, where the errors are within the width of lines.}
\end{figure}

Figure~\ref{fig:energy_beta_gamma} shows the $\beta$ dependence of $\varepsilon$ and $\rho$ for $\gamma=2.0$, $1.1$ and $0.9$ at mean degree $c=2.0$. 
The numerical results are calculated by the population-dynamics method with $10^5$ populations, and the statistical errors are as large as the width of each curve. 
As a property that should be satisfied in equilibrium, the energy density is a monotonically decreasing function of $\beta$, while $\rho$ and $\nu$ are not monotonically decreasing functions, as can be seen in Fig.~\ref{fig:energy_beta_gamma}.
For example, for $\gamma=1.1$, as $\beta$ is increased, $\rho$ is initially smaller than the minimum cover ratio before converging to the minimum cover ratio. 
This implies that $\beta$ should be larger to obtain the minimum cover ratio compared to the case with a larger $\gamma$, that is, $\gamma=2.0$. 
This is because the structure of excited states consisting of infeasible states depends on $\gamma$, and the number of low-lying excited states increases for smaller $\gamma$. 
When the system is restricted to only feasible states, the cover ratio decreases monotonically as $\varepsilon$ and $\rho$ coincide, and this non-monotonic behavior, due to the constraint relaxation, is eliminated.
For $\gamma<1$, that is, $\gamma=0.9$, $\varepsilon$ and $\rho$ converge to values lower than the minimum cover ratio.

\begin{figure}[h]
    \includegraphics[width=0.9\linewidth]{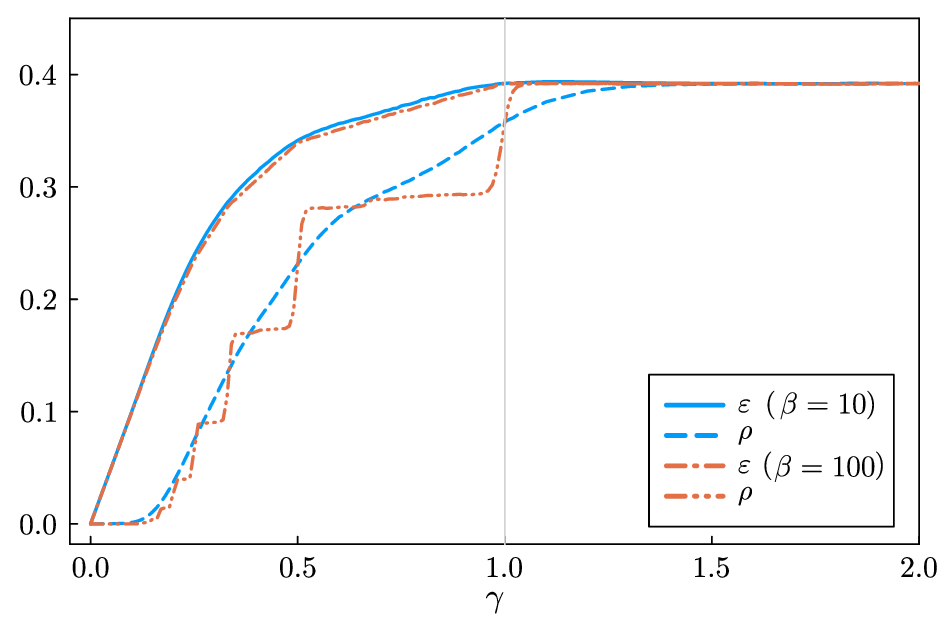}
    \caption{\label{fig:energy_gamma_at_large_beta}(Color Online) Penarty-coefficient $\gamma$ dependence of the energy density $\varepsilon$ and the cover ratio $\rho$ for $\beta=10$ (solid and dashed lines) and $100$ (dashed and two-dotted lines) at mean degree $c=2.0$. The simulation conditions of the population dynamics are the same as in Fig.~\ref{fig:energy_beta_gamma}. Since a large amount of data was taken in the area with finely varying curves, markers are not displayed for visibility.}
\end{figure}

The $\gamma$ dependence of $\varepsilon$ and $\rho$ with $c=2.0$ at low temperatures of $\beta=10$ and $100$ is shown in Fig.~\ref{fig:energy_gamma_at_large_beta}. 
For $\gamma$ greater than 1.5, $\varepsilon$ and $\rho$ are close to the minimum cover ratio at low temperatures. 
By contrast, in the $0<\gamma\le1$ region, the cover ratio changes abruptly in a staircase-like manner near rational numbers of $\gamma$, such as $1$, $1/2$, and $1/3$, and the mean energy, which is the sum of $\rho$ and $\nu$, changes slowly with respect to $\gamma$, depending on $\beta$.
This is a consequence of the exchange of cost and penalty contributions at the above points of $\gamma$, which will be discussed in detail as a ground-state property in the next section.  


\section{\label{sec:ground_state}Ground-state properties}
This section discusses the properties of the low-temperature limit, that is, the ground state of MVC. 
First, we construct the solution of the self-consistent equation for general $\gamma$ including $\gamma\le 1$, and show that $\gamma>1$ is a condition for obtaining feasible solutions as the ground states, consistent with previous studies~\cite{weigt2001minimal, zhang2009stability}. 
We then explore the behavior of the system in the infeasible region of $\gamma\le 1$, which is important for understanding some techniques for finding optimal solutions via infeasible states, such as the method of adaptive penalty function~\cite{coello2002theoretical, glover1993user}. 

\subsection{Saddle-point solution $p(h)$ with finite $\gamma$}
In the self-consistent equation of Eq.~(\ref{eq:self-consistent}), only $K(h;\beta,\gamma)$ of Eq.~(\ref{eq:Kofh}) depends explicitly on $\beta$ and $\gamma$, and $p(h)$ depends on them through $K(h;\beta,\gamma)$. 
Taking the limit $\beta\to\infty$ leads to 
\begin{equation}
  K(h;\infty,\gamma) = \begin{cases}
       \gamma & (\gamma \le h), \\
       h      & (0 < h < \gamma), \\
       0      & (h\le0).
      \end{cases}
      \label{eq:Kofh_infty}
\end{equation}
Note that the support for $p(h)$ is $h\le1$; when $\gamma\ge 1$, $K(h;\infty,\gamma)$ for $h\le1$ does not depend on $\gamma$, and therefore, $p(h)$ also does not depend on $\gamma$, including $\gamma\to\infty$.
Thus, $p(h)$ in the low-temperature limit for $\gamma\ge 1$ is equivalent to that at $\gamma=\infty$.
Additionally, we will show later that the penalty ratio is zero for $\gamma>1$.
Therefore, any result obtained in the low-temperature limit for $\gamma>1$ is exactly the same as in the previous study~\cite{weigt2001minimal}, which restricted the system to the feasible states.

To find solutions to the saddle-point equation for $p(h)$ in the low-temperature limit, the previous study assumes that the only possible values of $h$ taken in $p(h)$ are integers of $h\le1$~\cite{weigt2001minimal}.
This ``integer ansatz" has been reported in other combinatorial optimization problems~\cite{biroli2000variational,zdeborova2006number}.
We notice that this ansatz can at least be justified numerically from an analysis of the time evolution of population dynamics, and based on this, we can construct irrational and rational ansatz that can be applied to the infeasible region. 
See Appendix \ref{sec:PD_at_zero_temp} for details.

For $0<\gamma\le 1$, the possible values of $h$ in $p(h)$ are restricted to the sum of any integer and an integer multiple of $\gamma$ that is less than or equal to 1.
Formally, the solution can be described as
\begin{equation}
  p(h)=\sum_{l,l'\in \mathbb{Z},l'+\gamma l \ge0} r_{l,l'} \delta(h-1+l+\gamma l'),
  \label{eq:pofh_zero_T_solution_irrational}
\end{equation}
where $r_{l,l'}$ is a positive weight and normalized as $\sum_{l,l'}r_{i,i'}=1$. 
If $\gamma$ is an irrational number, it is difficult to find a specific formula for $r_{l,l'}$ that satisfies the saddle-point equation, and the numerical approximation of $p(h)$ is obtained using population-dynamics methods.
Moreover, if $\gamma$ is a rational number with $m$ as the denominator, $p(h)$ can be simplified to 
\begin{equation}
  p(h)=\sum_{l=0}^\infty r_{l,m} \delta\left(h-1+\frac{l}{m}\right). 
  \label{eq:pofh_zero_T_solution_rational}
\end{equation}
Particularly, for $\gamma=1/m$ with $m=1,2,\dots$, the coefficients $r_{l,m}$ can be written as 
\begin{equation}
  r_{l,m} = \frac{R_m^l}{l!} e^{-R_m} ,
\end{equation}
where $R_m$, depending on $c$, is a real solution of the equation, 
\begin{equation}
  R_m e^{R_m} = c\left(1 + R_m + \frac{R_m^2}{2!} + \cdots +  \frac{R_m^{m-1}}{(m-1)!}\right) .
\end{equation}
This formula is obtained by substituting Eq.~(\ref{eq:pofh_zero_T_solution_rational}) into Eq.~(\ref{eq:self-consistent}) and solving the simultaneous equations for $r_{l,m}$'s.
Note that for $m=1$, that is, $\gamma=1$, the solution $p(h)$ is exactly the same as that for $\gamma>1$ in the previous study~\cite{weigt2001minimal}.
Specifically, Eq.~(\ref{eq:pofh_zero_T_solution_rational}) represents the integer ansatz, and the formula reduces to the definition of Lambert's $W$ function~\cite{corless1996lambert}, $R_1(c)e^{R_1(c)}=c$, and $r_{l,1}=W(c)^{l+1}/cl!$.

Intuitively, these ansatzes can be considered a consequence of the discrete nature of the effective field distribution $p(h)$ in the low-temperature limit. 
For instance, for $\gamma > 1$, the energy gap from the ground state induced by a one-bit flip is determined by the number of uncovered vertices adjacent to each vertex, and the integer ansatz can be regarded as a property of the effective field of this discrete structure. 
In the case of $\gamma \le 1$, each vertex in the ground state can either satisfy or violate the constraint, and the energy gap takes various values composed of integers and integer multiples of $\gamma$ as shown in Eq.~(\ref{eq:pofh_zero_T_solution_irrational}). 
This intuitive understanding is consistent with the interpretation of the cover ratio and violation ratio in the low-temperature limit, as discussed in the next section.

\subsection{Nonbackbone and undetermined constraints}

Generally, there are multiple optimal solutions for a single instance $G$ of MCV.
Following Ref.~\cite{weigt2001minimal}, vertices that are commonly covered or uncovered by all solutions in common are both called ``backbone''. 
By contrast, vertices other than the backbone are called ``nonbackbone'', which are covered or uncovered by each optimal solution. 
From the definition of $p(h)$ in Eq.~(\ref{eqn:eff_potential}), the coefficients of $p(h)$ for $h<0$ in the ground state at $\beta\to\infty$ can be interpreted as the fraction of covered backbone and that for $h>0$ as the fraction of uncovered backbone. Moreover, that for $h=0$ is considered to be the fraction of nonbackbone~\cite{weigt2001minimal}. 

The cover ratio $\rho$ and penalty ratio $\nu$ in the low-temperature limit are obtained by taking $\beta\to\infty$ in Eq.~(\ref{eq:coverratio}) and (\ref{eq:penaltyratio}), respectively.
The low-temperature limit of $p(h)$ can be calculated by Eq.~(\ref{eq:self-consistent}) and (\ref{eq:Kofh_infty}) with the ansatz above.
Naively, $\rho$ and $\nu$ in these expressions are expected to be obtained by the integral of the product of the low-temperature limit of $p(h)$ and each factor in Eq.~(\ref{eq:coverratio}) and (\ref{eq:penaltyratio}), respectively.

The factor for the cover ratio in Eq.~(\ref{eq:coverratio}) reads 
\begin{equation}
 \lim_{\beta\to\infty}\frac{1}{1+e^{\beta h}}
  = \begin{cases}
   0 & (0<h) ,\\
   1/2 & (h=0) ,\\
   1 & (h<0) .
  \end{cases}
  \label{eq:min_cov_factor}
\end{equation}
This factor is constant for each interval of $h$, and the contribution of $p(h)$ to the cover ratio can be interpreted according to the previous study~\cite{weigt2001minimal} as follows. 
The contribution of the fraction of covered backbones $p(h)$ to the cover ratio is 1 when $h<0$. 
Moreover, the fraction of nonbackbones, $p(0)$, contributes 1/2 to the cover ratio, which assumes that the nonbackbone vertices are independently covered with a probability of 1/2 in the ground states.
As a result, when $\gamma\ge 1$, the cover ratio is independent of $\gamma$, and obtained as
\begin{equation}
 \rho(\infty,\gamma\ge1)
 = \frac{1}{2}r_{1,1}+\sum_{l=2}^\infty r_{l,1} 
 = 1 - \frac{W(c)}{c} - \frac{1}{2}\frac{W^2(c)}{c} ,
 \label{eq:min_cov_ratio}
\end{equation}
which coincides with the well-known minimum cover ratio, $x_c(c) = \rho(\infty,\gamma>1)$, under the RS assumption~\cite{weigt2001minimal}. 
This result can be modified by considering a correction field, which will be discussed in the next subsection.

Similarly, the factor of the penalty ratio in Eq.~(\ref{eq:penaltyratio}) in the low-temperature limit yields 
\begin{align}
  &\lim_{\beta\to\infty} \frac{1}{1+e^{-\beta (h-\gamma)}+e^{-\beta (h'-\gamma)}+e^{-\beta(h+h'-\gamma)}} \nonumber \\
  &= \begin{cases}
  1 & (h,h'>\gamma), \\
  1/2 & (h=\gamma,h'>\gamma \text{ or } h \leftrightarrow h'), \\
  1/3 & (h=h'=\gamma), \\
  0 & (h<\gamma \text{ or } h'<\gamma). 
  \end{cases}
  \label{eq:min_pen_factor}
\end{align}
The meaning of this factor is examined in detail below. 
When both $h$ and $h'$ are greater than $\gamma$ and the contribution is 1, the edge corresponds to a violated edge where both vertices on the edge are not covered in all ground states. 
When at least one of $h$ and $h'$ is smaller than $\gamma$ and the contribution is 0, it corresponds to an edge for which the constraint is always satisfied, that is, at least one of the vertices on the edge is covered.
In other words, edges with $h<\gamma$ and $h'>\gamma$ can be interpreted as each vertex on the edge being covered or uncovered, respectively.
By contrast, the vertices with $h=\gamma$ can be considered to be connected edges for which constraint satisfaction is undetermined.
Since the contribution of Eq.~(\ref{eq:min_pen_factor}) to the integral in Eq.~(\ref{eq:penaltyratio}) is 0 for $\gamma>1$, the above interpretation is possible only when $0\le\gamma\le 1$.

The value of $\nu$ is positive only if $\gamma\le 1$. 
For example, when $\gamma=1$, from the term $h=h'=\gamma$ in Eqs.~(\ref{eq:penaltyratio}) and (\ref{eq:min_pen_factor}), the penalty ratio yields 
\begin{equation}
    \nu(\infty,1) = \frac{c}{2}\frac{1}{3}r_{0,1}^2 = \frac{W^2(c)}{6c}.
    \label{eq:penalty_ratio_at_g1} 
\end{equation}
This result is inconsistent with the discussion of the penalty function in Sec.~\ref{Sec:MVCandPen} and numerically obtained $\varepsilon$ at sufficiently low temperature, for example, $\beta=100$, as shown in Fig.~\ref{fig:energy_gamma_at_large_beta}.
Since the feasible and infeasible ground states coexist just at $\gamma=1$, their ground energy must be equal to the ground energy of $\gamma>1$, and the numerical results approximate a smooth continuous function at $\gamma=1$.
However, in the above analytical calculation, $\varepsilon(\infty,1)=x_c(c)+\nu(\infty,1)$ and $\varepsilon(\infty,\gamma>1)=x_c(c)$, which is discontinuous at $\gamma=1$. 
The correction term to the effective field discussed in the next section also resolves the issue of this energy discontinuity.

\subsection{Correction field for undetermined vertices and edges}
\label{sec:correction_field}
A correction field called the `evanescent part' of the effective field $h$ has been introduced to calculate the contribution to entropy from the undetermined vertices of the ground states~\cite{biroli2000variational,zdeborova2006number,weigt2001minimal,zhang2009stability}. 
It is shown in the following that this correction field affects not only the entropy but also the energy density in MCV, particularly making the energy density continuous in the low-temperature limit and eliminating the discontinuity mentioned above.

The numerical observations by the population-dynamics method for $\beta\gg1$ show that $p(h)$ has a non-negligible finite width around each multiple sharp peak.
Let $\{h'\}$ be the set of these peak positions, for example, $h'=1,1-1/m,1-2/m,\dots$ for $\gamma=1/m$.
Then, the distribution of the effective field can be represented as the product of each delta peak and its surrounding distribution $\rho_{h'}(\tilde{h})$, expressed as
\begin{equation}
    p(h,\tilde{h}) = \sum_{\{h'\}} r_{h'}\delta(h-h')\rho_{h'}(\tilde{h}) ,
    \label{pofh_evanescent}
\end{equation}
where each $\rho_{h'}(\tilde{h})$ is assumed to be normalized in the $\beta\rightarrow 0$ limit. 
The one-variable distribution $p(h)$ can be reproduced by integrating over $\tilde{h}$ with a delta-function kernel as 
\begin{equation}
    p(h)=\int dh'd\tilde{h}\, \delta\!\left(h-h'-\frac{\tilde{h}}{\beta}\right)p(h',\tilde{h}).
    \label{eq:h_to_h_plus_h_tilde}
\end{equation}
Substituting Eq.~(\ref{eq:h_to_h_plus_h_tilde}) into Eq.~(\ref{eq:self-consistent}) and taking the limit of $\beta\to\infty$ with fixed $\gamma$, the self-consistent equation for $p(h,\tilde{h})$ reads
\begin{align}
    &p(h,\tilde{h}) = e^{-c}\sum_{l=0}^\infty \frac{c^l}{l!} 
  \int \prod_{i=1}^l dh_id\tilde{h}_i \, p(h_i, \tilde{h}_i) \nonumber\\
  &\times \delta\!\left(h - 1 + \sum_{j=1}^l K(h_j;\infty,\gamma) \right)
  \delta\!\left(\tilde{h} + \sum_{j=1}^l \tilde{K}(\tilde{h}_j \mid h_j)\right), 
  \label{eq:self-consistent-pofhh}
\end{align}
where the first delta function is the same as that of Eq.~(\ref{eq:self-consistent}) with $K$ in Eq.~(\ref{eq:Kofh_infty}).
Thus, the coefficients $r_{h'}$ of Eq.~(\ref{pofh_evanescent}) are the same as those of $p(h)$.
The update functions $ \tilde{K}(\tilde{h} \mid h)$ in the second delta function as a function $\tilde{h}$ conditional on $h$ is given by 
\begin{equation}
  \tilde{K}(\tilde{h} \mid h) \approx 
  \begin{cases}
    0 & (\gamma < h), \\
    - \log(1+e^{-\tilde{h}}) & (h=\gamma), \\
    \tilde{h} & (0<h<\gamma), \\
    \log(1+e^{\tilde{h}}) & (h=0), \\
    0 & (h<0).
  \end{cases}
  \label{eq:update_corr_field}
\end{equation} 
When $\gamma>1$, the solution $p(h,\tilde{h})$ of the self-consistent equation of Eq.~(\ref{eq:self-consistent-pofhh}) is mathematically equivalent to that in the previous studies~\cite{weigt2001minimal,zhang2009stability}.
See Appendix~\ref{sec:calc_corr_field} for detailed calculations.
The modified self-consistent equation is solved with the population-dynamics method on two sets of variables. 
See Appendix~\ref{sec:population_dynamics} for details. 

To obtain an expression for the cover ratio with the correction fields, substituting Eq.~(\ref{eq:h_to_h_plus_h_tilde}) into Eq.~(\ref{eq:coverratio}) and taking the limit of $\beta\to\infty$ yield the factor of the cover ratio as 
\begin{equation}
  \lim_{\beta\to\infty}\frac{1}{1+e^{\beta h + \tilde{h}}}
  = 
  \begin{cases}
    0 & (h>0) ,\\
    1/(1+e^{\tilde{h}}) & (h=0) ,\\
    1 & (h<1) .
  \end{cases}
  \label{eq:cov_factor_corrected}
\end{equation}
It is worth noting that the correction field only affects the nonbackbone term with $h=0$ and shifts its weight from 1/2 in Eq.~(\ref{eq:min_cov_factor}) which is the probability of nonbackbone vertices covered without a correction field. 
In fact, the cover ratio for $\gamma\ge 1$ in Eq.~(\ref{eq:min_cov_ratio}) is modified to
\begin{equation}
 \tilde{\rho}(\infty,\gamma\ge1)
 = 1 - \frac{W(c)}{c} + \frac{W^2(c)}{c} \left(\int d\tilde{h} \frac{\rho_0(\tilde{h})}{1+e^{\tilde{h}}} -1\right).
 \label{eq:correct_min_cov_ratio}
\end{equation}
The integral for $\tilde{h}$ in this formula is the probability of the nonbackbone being covered, 
which equals 1/2 in the case of \(\rho_0(\tilde{h}) = \rho_0(-\tilde{h})\), resulting in identical coefficients in Eq.~(\ref{eq:min_cov_ratio}) without correction.
Note that $\rho_0(\tilde{h})$ depends implicitly on $\gamma$ through Eq.~(\ref{eq:update_corr_field}) and converges to a different distribution in the low-temperature limit of $\gamma=1$ and $\gamma>1$.
Thus, the cover ratio $\tilde{\rho}(\infty,1)$ at $\gamma=1$ is different from that of $\gamma>1$. Here, the minimum cover ratio modified by the correction field, which depends on the mean degree $c$, is denoted by  $\tilde{x}_c(c) = \tilde{\rho}(\infty,\gamma>1)$.

For the penalty ratio, the correction field affects only the $h=\gamma$ term, which is the contribution from the vertices of edges for which constraint satisfaction is not determined. 
Since the constraints must be satisfied at all edges, the penalty ratio is 0 for $\gamma>1$, but the penalty ratio at $\gamma=1$ is positive to modify from Eq.~(\ref{eq:penalty_ratio_at_g1}) to
\begin{equation}
    \tilde{\nu}(\infty,1) =  \frac{W^2(c)}{2c} \int d\tilde{h}d\tilde{h}'\frac{\rho_1(\tilde{h})\rho_1(\tilde{h}')}{1+e^{-\tilde{h}}+e^{-\tilde{h}'}}.
    \label{eq:correct_penalty_ratio_at_g1}
\end{equation}
See also Appendix~\ref{sec:calc_corr_field} for the derivation.
Although it could not be shown analytically, we observed numerically that $\tilde{\rho}(\infty,1)+\tilde{\nu}(\infty,1)=\tilde{x}_c(c)$ holds in a wide range of $c$, that is, the ground-state energy density is equal for $\gamma=1$ and $\gamma>1$, and there is no energy discontinuity. 

\begin{figure}[t]
    \includegraphics[width=0.9\linewidth]{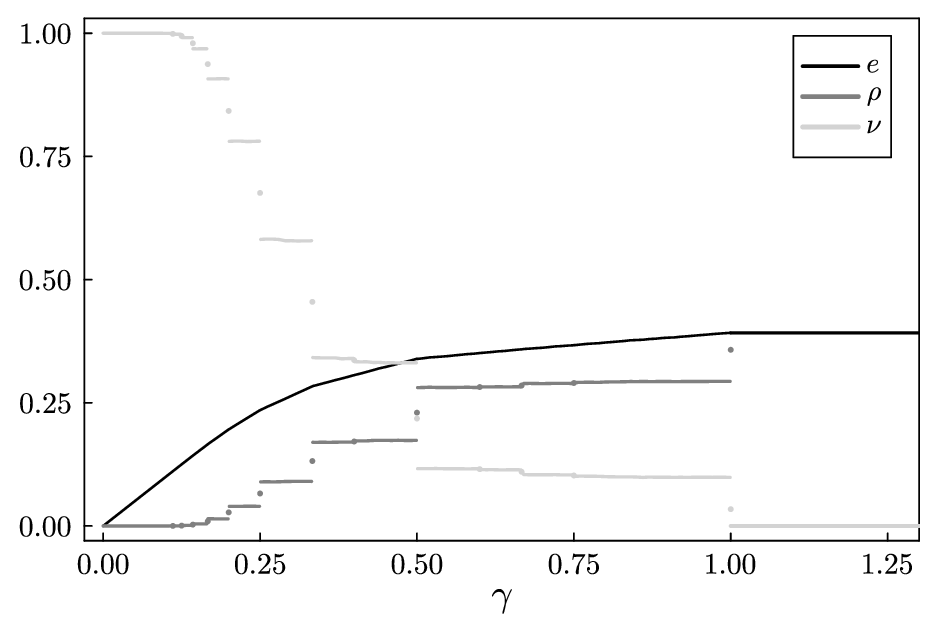}
    \caption{\label{fig:rho_nu_e_at_zero_T} $\gamma$ dependence of the cover ratio $\rho$, penalty ratio $\nu$ and energy density $\varepsilon$ in the low-temperature limit at mean degree $c=2.0$. The simulation conditions of population dynamics are the same as in Fig.~\ref{fig:energy_beta_gamma}, and the markers are not displayed for visibility.}
\end{figure}
Figure~\ref{fig:rho_nu_e_at_zero_T} shows the $\gamma$ dependence of $\varepsilon$, $\rho$ and $\nu$ in the low-temperature limit for $c=2.0$, obtained by solving the modified self-consistent equation, Eq.~(\ref{eq:self-consistent-pofhh}), using the population-dynamics method with $10^5$ population.
The energy density $\varepsilon$ increases monotonically and continuously with $\gamma$ from 0 to the minimum cover ratio $\tilde{x}_c(c)$.  
The cover ratio $\rho$ increases monotonically from 0 to $\tilde{x}_c(c)$, and the penalty ratio $\nu$ decreases monotonically from $c/2$ to 0.
In contrast to $\varepsilon$, the values of $\rho$ and $\nu$ show jumps at $\gamma=1$, $1/2$, $1/3$, $\dots$. 
Similar discontinuities are also observed at $\gamma=3/4$, $2/3$, etc., but with smaller amounts of jumps. 

This discontinuity in $\rho$ and $\nu$ can be explained by the energy function of the penalty method in Eq.~(\ref{eq:energywithpenalty}).
As discussed before, the condition for the coexistence of feasible and infeasible states is $\gamma=1$.
The feasible minimum-cover states are preferred at $\gamma=1+\delta$ with infinitesimal positive $\delta$. 
The infeasible states where a vertex is uncovered, and the violation is increased by $1$ are preferred at $\gamma=1-\delta$.
The amount of jumps is expected to be proportional to the fraction of such violated vertices.

A similar situation is observed at other rational $\gamma$.
Since there is no contribution from connected covered vertices, at $\gamma=1/m$, the amount of jump, which is the contribution of the fraction of a single covered vertex connected to $m$ uncovered vertices, is large.
The jump is small for the other $\gamma$ values that require multiple covered vertices to be involved, such as 3/4 or 2/3. 
Since any combination of vertices and edges can exist in the thermodynamic limit, $\rho$ and $\nu$ can be discontinuous at $\gamma$ of all rational numbers, resulting in a devil's staircase-like structure.
While not shown here, it should be noted that numerical experiments using a generated instance also observed these significant jumps in $\rho$ and $\nu$ at $1/m$.

\begin{figure}[t]
    \includegraphics[width=0.9\linewidth]{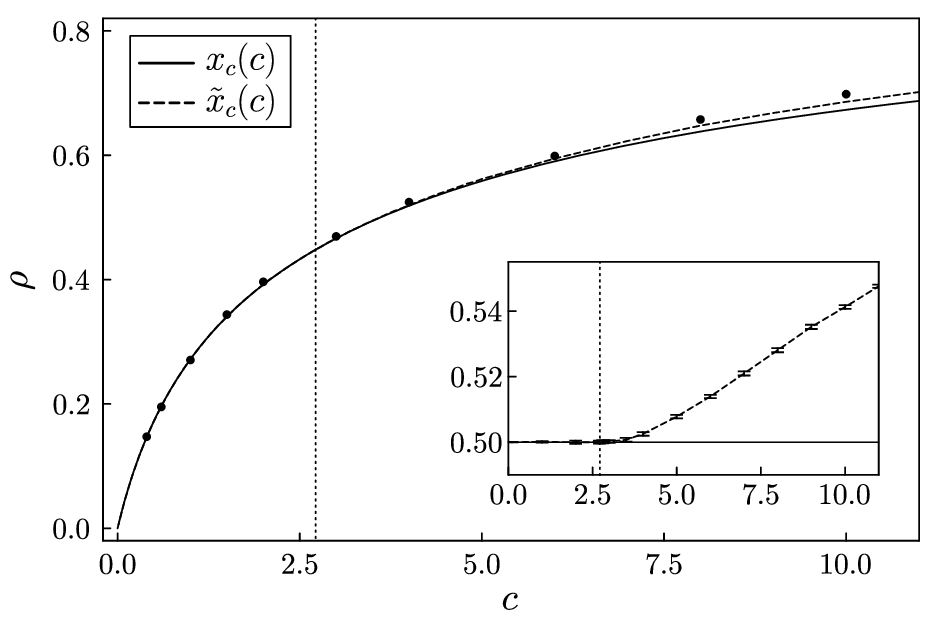}
    \caption{\label{fig:min_coverratio}$c$ dependence of the minimum cover ratios, $x_c(c)$ and $\tilde{x}_c(c)$ without and with the correction field (solid and dashed lines), respectively. Small circles denote the Monte Carlo result obtained by~\cite{weigt2001minimal}. The vertical dotted line denotes the known RS/RSB boundary of $c=e$. 
    The inset shows $c$ dependence of the probability of nonbackbone vertices being covered, which is 1/2 without the correction field.}
\end{figure}
The correction field discussed here also provides a small but significant correction to the cover ratio of the RS ansatz.
The $c$-dependence of the minimum cover ratio is shown in Fig.~\ref{fig:min_coverratio}, as well as the previous results obtained by the Monte Carlo method~\cite{weigt2001minimal}. It is shown that the correction term in the effective field introduced slightly improves the approximation accuracy of the RS solution. 
The inset of Fig.~\ref{fig:min_coverratio} shows the probability of a nonbackbone vertex being covered.
The contribution of the correction term completely disappears in the RS region of $c<e$, and is truly larger than 1/2 for $c > e$. 
Although our correction term is still under the RS assumption, it is interesting to note that the effect of the correction term appears only in the RSB region. 

\begin{figure}[t]
    \includegraphics[width=0.9\linewidth]{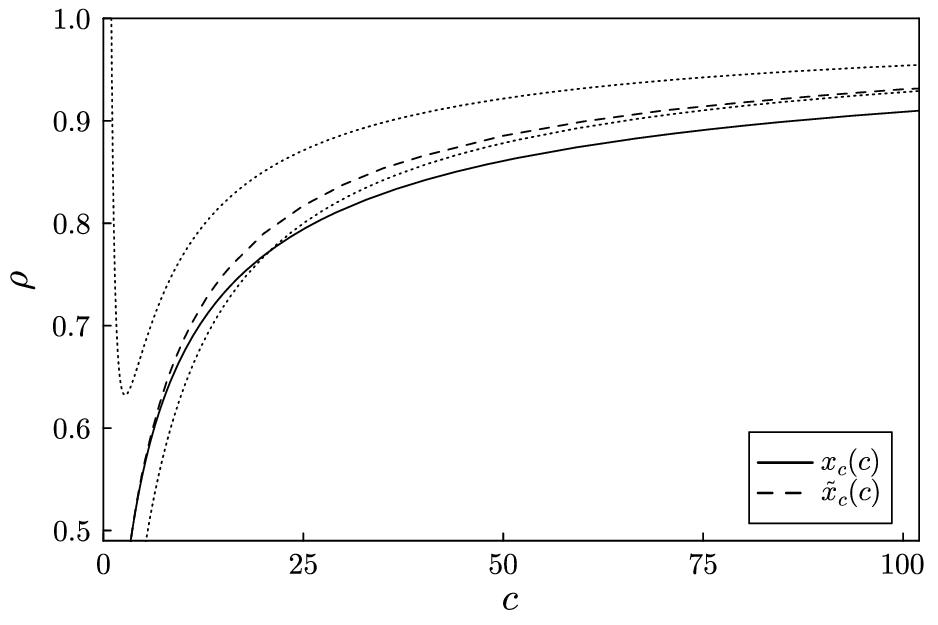}
    \caption{\label{fig:min_vc_bounds}$c$ dependence of the minimum cover ratios, $x_c(c)$ and $\tilde{x}_c(c)$ without and with correction field (solid and dashed lines), respectively.
    Dotted lines denote upper and lower bounds in~\cite{gazmuri1984independent}.} 
\end{figure}

Another notable observation is that as $c$ is increased, $\tilde{x}_c(c)$ approaches the lower bound obtained by the combinatorial analysis~\cite{gazmuri1984independent}. 
The bound is given by $x_l(c)<x_c(c)<1-\ln{c}/c$ for $c\ge 1$, where the lower bound $x_l(c)$ is the solution of $x_l(c)\ln{ x_l(c)}+(1-x_l(c))\ln(1-x_l(c))+(c/2)(1-x_l(c))^2=0$.
In Fig.~\ref{fig:min_vc_bounds}, the original RS estimation, $x_c(c)$,  violates the lower bound for $c\gtrsim20.7$, but the modified estimation, $\tilde{x}_c(c)$, is truly greater than the lower bound within statistical error and appears asymptotically coincide with the bound for sufficiently large $c$. 

\section{Stability analysis of self-consistent equation} \label{sec:stabilityofselfconsistent}
In this section, we consider the self-consistent equation for $p(h)$ as an iterative equation and discuss the linear stability of its solution $p(h)$.
This analysis corresponds to determining convergence conditions when solving the iterative equation using the population dynamics method. 
In the first part of this section, we examine stability boundaries in the low-temperature limit that can be analytically analyzed. 
In the latter part, we explore the stability at finite temperatures, demonstrating that these boundaries coincide with the conditions for $p(h)$ to appear as an oscillatory solution. 
Furthermore, we analyze the effect of damping, a method to suppress oscillatory solutions, using the approach here.

The self-consistent equation for $p(h)$ in Eq.~(\ref{eq:self-consistent}) is the functional equation for $p(h)$ with $K(h;\beta,\gamma)$ as the parameter function.
With the functional operator $P_{\rm SC}$ on the right-hand side of Eq.~(\ref{eq:self-consistent}), the functional equation is expressed as
\begin{equation}
    p(h) = P_{\rm SC}[p(h)] .
    \label{eqn:func_eq}
\end{equation}
One would naively assume that local stability is required for the equation to be solvable by a forward iterative method, one of the methods for solving the self-consistent equation. 
Suppose $p(h)$ is a solution to this equation and $\delta p(h)$ is a variation around the solution $p(h)$;  
the normalization condition of $p(h)$ imposes on $\delta p(h)$ the condition $\int dh \delta p(h)=0$. 
Substituting this into Eq.~(\ref{eq:self-consistent}) and considering up to the first-order variation, the equation reads
\begin{align}
  \delta p(h) = c \int dh' p(h+K(h';\beta,\gamma)) \delta p(h') .
  \label{eq:eigenvalue_equation_for_delta_p}
\end{align}
Note that the derivation does not depend on the specific form of $p(h)$ and $K(h;\beta,\gamma)$, but only on the factor of the Poisson distribution in Eq.~(\ref{eq:self-consistent}), thus this equation can be applied to other combinatorial problems defined on ER random graphs.
Suppose the right-hand side of Eq.~(\ref{eq:eigenvalue_equation_for_delta_p}) is the linear integral operator $\hat{P}$, then $\delta p(h)$ is an eigenfunction of $\hat{P}$, that satisfies the equation.  
\begin{equation}
    \delta p(h) = c\hat{P} [\delta p(h)] .
    \label{eq:eigenvalue_eq}
\end{equation}

Since $p(h+K(h'))$ is a stochastic kernel, it follows that the largest eigenvalue of $\hat{P}$ is 1, and the absolute values of other eigenvalues are less than 1.
The eigenfunctions belonging to eigenvalues less than 1 are orthogonal to the constant function, that is, their integral over $h$ is zero.
Let the eigenvalues of $\hat{P}$ be in descending order of absolute value, that is, $1=|\lambda_0|>|\lambda_1|\ge\cdots$, and let the corresponding eigenfunctions be $v_0(h),v_1(h),\ldots$, which satisfy $\hat{P} v_i(h)=\lambda_iv_i(h)$ and $\int dh v_i(h) = 0$ for $i\ge1$.
Thus, any variational $\delta p_0(h)$ can be expanded by the eigenfunctions of $\hat{P}$ except $v_0(h)$, as
\begin{equation}
  \delta p(h)= \sum_{i=1}^\infty a_i v_i(h), 
\end{equation}
where $a_i$ is the expansion coefficient. 
Then, the local stability condition for the solution is that the second eigenvalue of $\hat{P}$ satisfies 
\begin{equation}
  c|\lambda_1|<1 .
  \label{eqn:instability_condition}
\end{equation}
It is expected that when this condition is satisfied, the solution $p(h)$ of the self-consistent equation is solved by the iterative method.

\subsection{Zero temperature limit}
Here, we discuss the cases $\gamma>1$ and $\gamma=1/m$ in the low-temperature limit by analytically obtaining the eigenvalues.
For $\gamma\ge1$, $p(h)$ at $\beta=\infty$ consists of the sum of delta functions with peaks at integers $h=1-l$ for all natural numbers $l$; thus Eq.~(\ref{eq:eigenvalue_eq}) is an eigenvalue equation of the infinite order matrix with the coefficients $r_l=r_{l,1}$, which can be given analytically.
Let $P_l$ be the first $(l+1)\times (l+1)$ submatrix of the infinite matrix $\hat{P}$, then it is given by 
\begin{equation}
  P_l = \begin{pmatrix}
        0 & r_0 & r_0 & \cdots & r_0 \\
        r_0 & r_1 & r_1 & \cdots & r_1 \\
        r_1 & r_2 & r_2 & \cdots & r_2 \\
        \vdots & \vdots & \vdots & \ddots & \vdots\\
        r_{l-1} & r_l & r_l & \cdots & r_l
      \end{pmatrix}.
\end{equation}
Assuming $\sum_l^\infty r_l=1$ and $r_l\to 0$ when $l\to\infty$, the eigenvalues of $\hat{P}$ are $\lambda=1,-r_0$ and $0$ in decreasing order of the absolute value, with the eigenvalues $0$ degenerating to infinity.
The convergence condition is $cr_0=W(c)<1$, resulting in the well-known stability limit of the RS solution at the low-temperature limit, $c=e$~\cite{weigt2001minimal}.

For $\gamma=1/m$, the corresponding matrix $P_{l,m}$ is also defined from Eq.~(\ref{eq:pofh_zero_T_solution_rational}) as 
\begin{equation}
    P_{l,m} = \begin{pmatrix}
        0 & \cdots & 0 & r_{0,m} & \cdots & r_{0,m} \\ 
        r_{0,m} & \cdots & r_{0,m} & r_{1,m} & \cdots & r_{1,m} \\ 
        \vdots &  & \vdots & \vdots & & \vdots \\
        r_{l-1,m} & \cdots & r_{l-1,m} & r_{l,m} & \cdots & r_{l,m}
    \end{pmatrix}, 
\end{equation}
where the first $m$ columns of the $l+1$ square matrix are shifted down by one. 
The eigenvalues for $l\to\infty$ are given by 
\begin{equation}
    \lambda = 1, -r_{m-1,m}, 0,
\end{equation}
in decreasing order of absolute value. 
The condition for $p(h)$ to converge is $cr_{m-1}<1$, and we can find the maximum value of $c$ at which $p(h)$ converges for any $\gamma=1/m$. 

\begin{figure}[t]
    \includegraphics[width=0.9\linewidth]{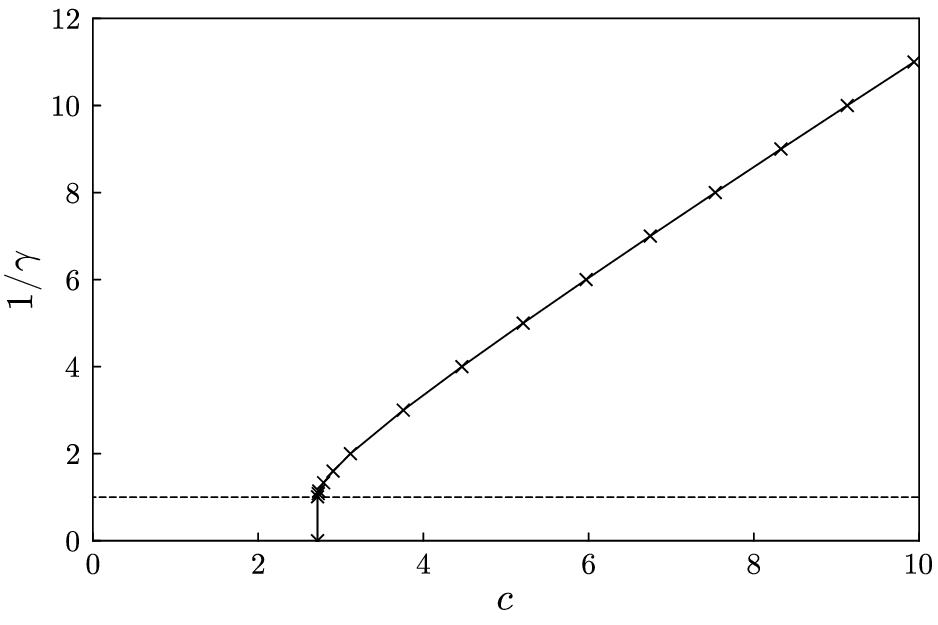}
    \caption{\label{fig:zerotempstabilitybound} Stability bound for the self-consistent equation of Eq.~(\ref{eq:self-consistent}) in the low-temperature limit as a function of mean degree $c$, above which the stability condition is satisfied. The dashed line is the limit of the RS solution that can be valid for $\gamma\to\infty$~\cite{zhang2009stability}, which will be discussed in the next section.}
\end{figure}
It is also possible to determine the eigenvalues of the matrix $\hat{P}$ for any rational number $\gamma<1$ by estimating the matrix elements using the population dynamics method.
Figure~\ref{fig:zerotempstabilitybound} shows the stability bounds of $1/\gamma$ for each value of $c$ in the low-temperature limit evaluated by the population dynamics method with $10^5$ populations. 
When the system is restricted to the feasible state at $1/\gamma=0$, the stable region coincides with the RS stable region, $c<e$~\cite{weigt2001minimal}, and this bound does not change in the range $1/\gamma<1$. 
When the constraint of the system is relaxed to the infeasible region for $1/\gamma>1$, the stability bound for $p(h)$ is extended to $c>e$.
This bound is asymptotically consistent with the instability conditions of the RS solution discussed in the next section and is expected to indicate the correct phase boundary.

\subsection{\label{sec:damping}Finite temperature and damping}
For finite $\beta$, the integral equation, Eq.~(\ref{eq:eigenvalue_equation_for_delta_p}) can be approximately solved by using the Fredholm method~\cite{courant2008methods}. 
Specifically, the integral equation can be reduced to an eigenvalue problem by obtaining $p(h)$ using the population dynamics method and discretizing the integral with respect to $h$ into the appropriate interval $\Delta h$. Eventually, the eigenvalues of Eq.~(\ref{eq:eigenvalue_equation_for_delta_p}) are obtained by numerically extrapolating the limit of $\Delta h\to0$ and the stability bound, Eq.~(\ref{eqn:instability_condition}), is evaluated from the second eigenvalue. 

\begin{figure}[t]
    \includegraphics[width=0.9\linewidth]{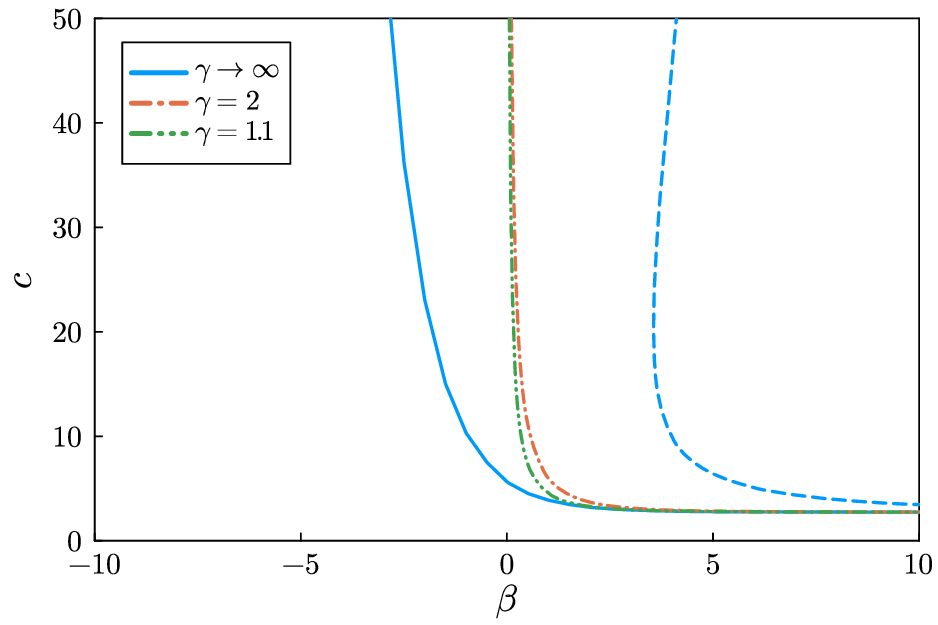}
    \caption{\label{fig:stability_pofh}(Color Online) Stability bounds for the self-consistent equation~(\ref{eq:self-consistent}) without damping at finite $\beta$ and mean degree $c$. The dashed line is the stability bound of the RS solution at $\gamma\to\infty$ obtained by the cavity method~\cite{zhang2009stability}.
    }
\end{figure}

Figure~\ref{fig:stability_pofh} shows the stability bounds in the plane of $\beta$ and $c$, for the case $\gamma\to\infty$, which restricts the system to the feasible states, and for the cases $\gamma=2.0$ and $1.1$. 
Decreasing $\gamma$ reduces the region where the iterative equations are stable, but all bounds converge to $c=e$ in the low-temperature limit. 

In fact, this bound is in good agreement with the bound where an oscillating solution appears when solved for the population dynamics method, indicating that this analysis correctly leads to the stability bound for the iterative method. 
When $c|\lambda_1|>1$, the amplitude of $\delta p(h)$ grows exponentially with the number of iterations, but the distribution eventually oscillates between the two distributions because of the negative eigenvalue. 
However, as shown in Fig.~\ref{fig:stability_pofh}, the boundary obtained by this method is different from the local stability bounds for $\gamma\to\infty$ in the cavity method, as discussed in the next section. Since no physical singularities are observed around this boundary, it is considered to be only a numerical technical problem not related to physical phenomena such as phase transitions. 

A method called damping is a heuristic often used to avoid oscillatory solutions in iterative methods such as population dynamics.
The update equation in the damping with the parameter $r$ 
is expressed as
\begin{equation}
    P_{\rm SC}[p_0(h);r] = (1-r)p_0(h) + r P_{\rm SC}[p_0(h)] .
\end{equation}
This shows the stability condition is modified to $\min_{i \ge 1}{|1-r+r c\lambda_i|} < 1$. 
Our numerical observation indicates that $\lambda_1$ may be negative for any $\beta$ and $\gamma$, in which case the convergence region can be expanded to some extent. 
In this sense, the stability conditions that depend explicitly on the parameter $r$ should not be related to any physical phenomenon. 
By contrast, the fact that only the zero-temperature limit is consistent with the RS/RSB transition may have some intrinsic meaning.

\section{stability analysis using the cavity method} \label{sec:stabilitybycavity}
In the previous section, we discussed the stability of the self-consistent equations, but its stability condition could not reproduce the phase transition boundary derived in the previous study~\cite{zhang2009stability} in the case of $\gamma=\infty$, except for the low-temperature limit. 
To study the phase transition of this system, including the case where $\gamma$ is finite, we use the cavity method of the system with the penalty function in this section. 
More specifically, we follow the method in Refs.~\cite{zdeborova2006number,zhang2009stability} to investigate the stability limit of the RS solution by the divergence of the spin-glass susceptibility.

\subsection{Replica symmetric cavity method}
In this subsection, we derive the belief propagation (BP) equation for MVC with the penalty function using the cavity method as outlined in Ref.~\cite{mezard2001bethe}.
Given a graph $G(V, E)$, consider a ``cavity'' graph defined by removing one edge from the graph $G$. 
Let $P_{j \to i}(x_j)$ be the probability of variable $x_j$ at vertex $j$ after removing the edge $j\to i$, 
and $\partial j\setminus i$ be the set of vertices adjacent to vertex $j$ excluding $i$.
Assuming that the correlation between $x_k$ and $x_{k'}$ on different vertices $k,k'\in\partial j\setminus i$ is negligible, that is, $G$ is locally a tree, the probabilities $P_{k \to j}(x_k)$ follow the BP equation expressed as 
\begin{equation*}
    P_{j \to i}(x_j)
      = \frac{1}{Z_{j \to i}} \phi_j(x_j) \prod_{k \in \partial j\setminus i}  \sum_{x_k} \psi_{jk}(x_j, x_k) P_{k \to j}(x_k) ,
\end{equation*}
where $\phi_i(x_i)=e^{-\beta x_i}$ and $\psi_{ij}(x_i, x_j)=\exp(-\beta\gamma(1-x_i)(1-x_j))$ are the weights of vertex $i$ and edge $(ij)\in E$, respectively, and $Z_{j \to i}$ is a normalization constant.
Defining the cavity field $h_{j \to i}$ as $e^{-\beta h_{j \to i}} = P_{j \to i}(1)/P_{j \to i}(0)$, the BP equation using the cavity field for MVC with the penalty function reads
\begin{equation}
  h_{j \to i} = 1 - \sum_{k\in\partial j\setminus i} K(h_{k \to j};\beta,\gamma) ,
  \label{eq:cavity-eq}
\end{equation}
where $K(h;\beta,\gamma)$ is the same as that in Eq.~(\ref{eq:Kofh}). 
The limit $\gamma\to\infty$ yields an expression equivalent to previous studies where the system is restricted to the feasible states~\cite{zhou2003vertex,zhang2009stability}.

Equation~(\ref{eq:cavity-eq}) is a recursive equation on a given $G$.
The RS approximation in the cavity method requires that each cavity field is independently and identically distributed according to the probability distribution $p(h)$ for the cavity field.
Assuming that $G$ is an instance of the ER random graph with mean degree $c$ given by Eq.~(\ref{eq:ER}), for $N\to\infty$, the recursive equation of $p(h)$ reads
\begin{equation}
  p(h) = 
  e^{-c} \sum_{k=1}^\infty \frac{kc^k}{ck!} \int \prod_{i=1}^{k-1} dh_ip(h_i)\, 
  \delta\!\!\left(h-1+\sum_{j=1}^{k-1} K(h_j)\right). 
  \label{eq:self-consistent_cavity}
\end{equation}
where $e^{-c}kc^k /ck!$ is the edge-perspective degree distribution of the ER random graph and represents the distribution that $|\partial j|$ follows when an edge $j\to i$ is randomly chosen.
Substituting $k-1$ as $l$, this equation coincides with Eq.~(\ref{eq:self-consistent}), that is, the self-consistent equation of the effective field distribution in the replica method.

\subsection{Divergence of the spin-glass susceptibility}\label{sec:div_of_chiSG}

Following Refs.~\cite{zdeborova2006number,zhang2009stability}, we detect the boundary where the RS solution fails at finite temperature $\beta$ and constraint strength $\gamma$ via the divergence of the spin-glass susceptibility.
The spin-glass susceptibility of MVC is defined as
\begin{equation}
    \chi_{\rm SG} 
    = \frac{4^2}{N}\sum_{i\ne j}\langle x_ix_j\rangle_c^2 ,
\end{equation}
where $\langle x_ix_j\rangle_c$ is the connected correlation function between $x_i$ and $x_j$.
Decomposing this sum by the shortest distance $d$ between the two connected vertices $i$ and $j$, we get 
\begin{equation}
    \chi_{\rm SG} = \frac{4^2}{N} \sum_{i=1}^{N} \sum_{d=1}^N    \sum_{j\in i(d)}^{|i(d)|} \langle x_i x_j\rangle_c^2 ,
\end{equation}
where $i(d)$ is the set of all vertices connected by the shortest path with $d$ edges from the vertex $i$.

Assuming that the graph $G$ is regarded as a local tree, the fluctuation-dissipation theorem~\cite{zdeborova2006number} leads to the correlation function between $x_0$ and $x_d$ as 
\begin{align}
  \langle x_0 x_d \rangle_c^2 \propto \prod_{k=1}^{d-1} \left| \frac{\partial h_{k\to k+1}}{\partial h_{k-1\to k}} \right|^2
  = \prod_{k=1}^{d-1} \left| \frac{\partial K(h_{k-1\to k})}{\partial h_{k-1\to k}} \right|^2,
  \label{eq:corrfunc0d}
\end{align}
where the index $k$ denotes the vertices along the shortest path from $x_0$ to $x_d$, taken from $1$ to $d-1$.
The second equality follows from Eq.~(\ref{eq:cavity-eq}).
It is challenging to evaluate $\chi_{\rm SG}$ directly, except in special cases.
Instead, we define $\Delta_{j\to i}$ as the contribution of the overall shortest paths up to the length $N$ to each end of the tree (leaf), with edge $j\to i$ as the root to the right-hand side of Eq.~(\ref{eq:corrfunc0d}).  
This quantity has the same order of $\chi_{\rm SG}$ for $N$ and satisfies the following recursive equation:
\begin{equation}
    \Delta_{j\to i} = \sum_{k\in \partial j \setminus i} \left|\frac{\partial K(h_{k\to j};\beta,\gamma)}{\partial h_{k\to j}}\right|^2 \Delta_{k\to j},
    \label{eq:update_Delta}
\end{equation}
where $\{h_{j\to i}\}$ follows the recursive equation of Eq.~(\ref{eq:cavity-eq}) and the solution can be obtained simultaneously in an iterative manner. 

Applying the RS approximation to MVCs on the ER random graph leads to a recursive equation for the bivariate distribution, $p(h,\Delta)$, similar to Eq.~(\ref{eq:self-consistent-pofhh}). 
It should be noted that although the correlation among the individual edges can be ignored by the RS approximation, $p(h,\Delta)$ is a simultaneous distribution, and each sample in the population dynamics method must be treated as a paired variable. See Appendix \ref{sec:population_dynamics} for details on the computational method. 
The average value of $\Delta$ with respect to the population exceeds 1 after a sufficiently large number of iterations, meaning that $\chi_{\rm SG}$ diverges~\cite{zhang2009stability}.
In our numerical experiments, after the initial burn-in period, the maximum or mean value of $\Delta$ shows either a monotonically increasing or decreasing trend as the number of iterations progresses.
This trend allows us to detect the divergence of $\chi_{\rm SG}$.
\begin{figure}[t]
    \includegraphics[width=0.9\linewidth]{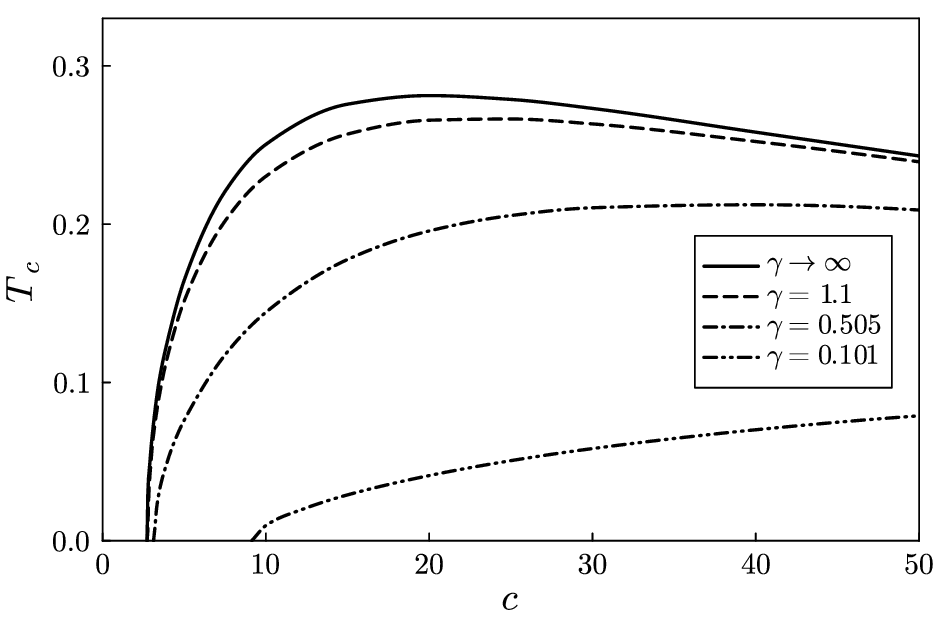}
    \caption{\label{fig:stability_c_T}Mean-degree $c$ dependence of the divergence temperature $T_{\rm c}$ of $\chi_{\rm SG}$ for several $\gamma$'s, obtained by the population dynamics methods. 
    Numerical errors are similar to the width of curves, and markers are omitted for visibility.
    }
\end{figure}

Figure~\ref{fig:stability_c_T} shows the $c$ dependence of the transition temperature $T_{\rm c}$ at which $\chi_{\rm SG}$ diverges at some $\gamma$ values.
This result is consistent with the previous study~\cite{zhang2009stability} in the limit of $\gamma\to\infty$. 
When decreasing $\gamma$ from infinity, the phase boundary shifts towards the low-temperature side.
Additionally, $T_\mathrm{c}$ shows a peak as a function of $c$, and reducing the constraints results in this peak shifting towards larger $c$ values.
Focusing on $T=0$, the phase boundary is $c=e$ independent of $\gamma$ in the range $\gamma>1$, but it shifts to the range $c>e$ for $\gamma<1$. 
These results are consistent with the stability analysis of the self-consistent equation, as shown in Fig.~\ref{fig:zerotempstabilitybound}. 

\begin{figure}[t]
    \includegraphics[width=0.9\linewidth]{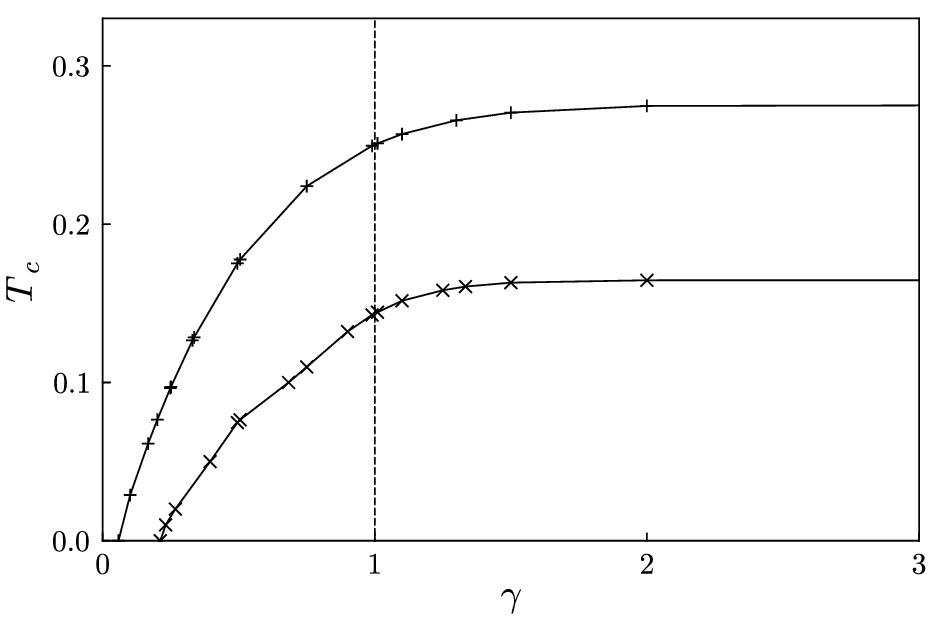}
    \caption{\label{fig:stability_gamma_T_c5}$\gamma$ dependence of the divergence temperature $T_{\rm c}$ of $\chi_{\rm SG}$ for $c=5.0$ (cross marker) and $15$ (plus). The vertical line represents the boundary between the feasible and infeasible regions at $T=0$. 
    Numerical errors are within the width of the curves.
    }
\end{figure}

Figure~\ref{fig:stability_gamma_T_c5} shows the $\gamma$ dependence of $T_{\rm c}$ for $c=5$ and 15. 
When the value of $\gamma$ is higher than $2$, the transition temperature $T_{\rm c}$ is approximately equal to that at $\gamma\to\infty$, and $T_{\rm c}$ decreases monotonically with decreasing $\gamma$. 
Eventually, for a finite value of $\gamma$, $T_{\rm c}$ converges to zero. 
This value is approximately $0.21$ for $c=5$ and $0.058$ for $c=15$.
In the limit $\gamma=0$, the system is equivalent to an independent Ising spin system under a uniform magnetic field, and the RS solution is always stable at any temperature. 

\section{MCMC Results} \label{sec:numericalresults}

In this section, we perform MCMC calculations for a finite-size MVC to verify the results assuming the RS discussed above to explore phenomena not captured by the RS analysis. 
The critical temperature $T_{\mathrm{c}}$ is estimated using finite-size scaling analysis, and the results are compared with those obtained using the cavity method described in the preceding section.
Furthermore, the RS approximation of the MVC is expected to fail at the transition temperatures, $T_{\mathrm{c}}$.
Previous numerical studies have revealed that the ground states of the MVC can be divided into different clusters~\cite{barthel2004clustering}, and we investigate the connection to it from finite temperatures.

\subsection{Numerical method and observables}
\label{sec:method_and_observables}

For the numerical experiments, we use the exchange Monte Carlo method (parallel tempering)~\cite{hukushima1996exchange}, which is effective for sampling from a state space separated into numerous subspaces.
In this method, $R$ independent copies (replicas) of MVC defined on the same instance $G$ undergo MCMC simulations independently at different inverse temperatures, such as $\beta_1<\beta_2<\dots<\beta_R$. 
The variables for each replica are updated using the single-bit-flip Metropolis algorithm~\cite{metropolis1953equation}, and $N$ local trials are called one Monte Carlo step (MCS). 
The number of replicas is set to $R=60$, and the highest and lowest temperatures are set to $\beta_1=0.1$ and $\beta_R=10$, respectively.
The temperature interval is adjusted so that the product $\Delta\beta\langle\Delta E\rangle$ of the temperature and energy difference between adjacent replicas is constant. 
Then, temperature swaps are executed every 1MCS between these replicas during MCMC simulations. 
While the correlation between samples in MCMC methods is a factor that reduces statistical accuracy, the longest correlation time scale in this method is considered to be the round-trip time $\tau_{\rm RT}$, which is the time for one replica to travel both the highest and the lowest temperatures and return to the original temperature~\cite{hukushima1996exchange}. 
For the calculations in this study, the sampling intervals were set to exceed $\tau_{\rm RT}/ R$. 
The round-trip time depends on $\gamma$, and $c$ particularly tends to increase with increasing $\gamma$. 
The specific value of the sampling interval was then at least 150 MCS, and sometimes greater than 600 MCS when $N$ was large. 
We also generated $500-100$ ER graphs for each size of $N=64$, $128$, $256$, $512$, and $1024$ to evaluate random graph averages.

The overlap distribution for a fixed graph $G$ is defined as
\begin{equation}
    P_G(q)=\langle \delta(q-q^{ab}) \rangle, 
\end{equation}
where $q^{ab}$ is the overlap between two states $\bm{x}^{(a)}$ and $\bm{x}^{(b)}$, defined as
\begin{equation}
    q^{ab} = \frac{1}{N}\sum_{i=1}^N (1-2x_i^{(a)})(1-2x_i^{(b)}).
\end{equation}
The overlap takes 1 for $\bm{x}^{(a)}=\bm{x}^{(b)}$ and $-1$ for $\bm{x}^{(a)}=\bm{1}-\bm{x}^{(b)}$, and is related to the Hamming distance density to $d_{\rm H}^{ab}=\sum_i(x_i^{(a)}-x_i^{(a)})^2/N = (1-q^{ab})/2$.
In the numerical experiment, we run two independent simulations and calculate $q^{ab}$ from two states at the same $\beta$ after a burn-in period to obtain $P_G(q)$ as the histogram over time. 
The overlap distribution is defined as the random-graph average of $P_G(q)$, given by 
\begin{equation}
    P(q) = [P_G(q)]_G .
\end{equation}
In the case of RS, this distribution is expected to be Gaussian in finite-size systems and the delta function, which is a trivial distribution, for an infinite system. 
However, in the region of RSB, it is expected to be nontrivial, that is, deviating from the delta-function-type distribution~\cite{parisi1999continuous}. The spin-glass susceptibility is obtained by the second moment of this distribution multiplied by $N$. 

\subsection{Finite-size scaling analysis of spin-glass susceptibility}

First, we performed a finite-size scaling analysis of $\chi_{\rm SG}$ obtained by the MCMC methods near the critical temperature $\beta_{\mathrm{c}}$ to examine the consistency of the transition temperature estimated by the cavity method.
In the finite-size scaling analysis, we assume that the scaling function can be expressed near $T_{\mathrm{c}}$ as
\begin{equation}
    \chi_{\rm SG}(N,T) = N^{x/y} f((T-T_\mathrm{c})N^{1/y}),
\end{equation}
where $f$ is the scaling function, $x$ is the critical exponent of $\chi_{\rm SG}$, and $y$ is the critical exponent with respect to the correlation size of the system.

\begin{figure}[t]
    \includegraphics[width=0.9\linewidth]{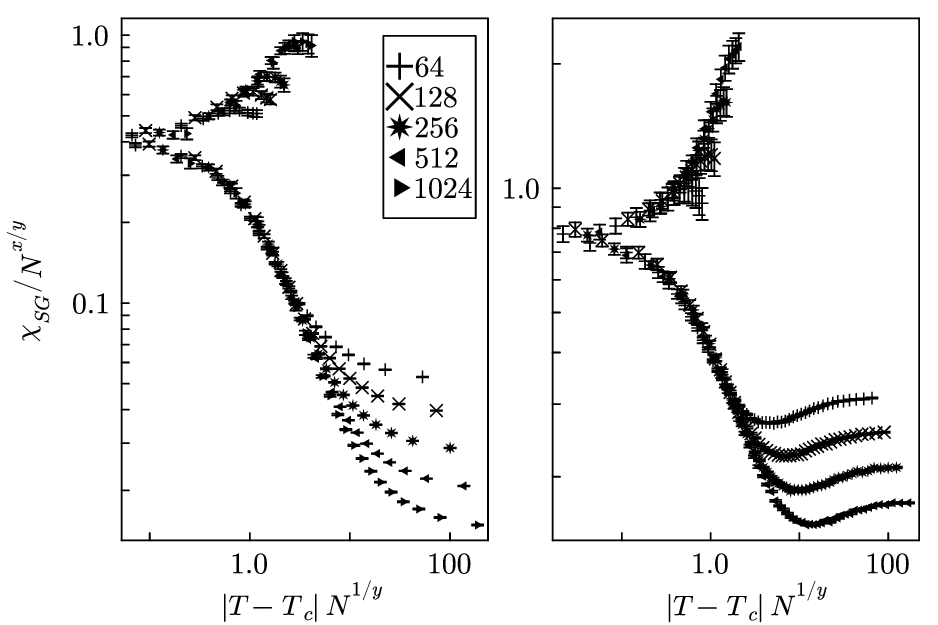}
    \caption{\label{fig:finite_size_scaling}Finite size scaling plots of the spin-glass susceptibility at $\gamma\to\infty$ (left) and $\gamma=0.505$ (right) for ER random graph with mean degree $c=15$. The left and right plots are obtained with $T_\mathrm{c}=0.275$ and $0.219$, and with critical exponents as $(x,y)=(1.06(2), 2.19(4))$ and $(0.61(4), 2.20(5))$, respectively.
    }
\end{figure}

Figure~\ref{fig:finite_size_scaling} shows the finite-size scaling plots with $\gamma=\infty$ and $0.505$ for $c=15$, where the transition temperatures $T_\mathrm{c}=0.275$ and $0.219$, respectively, are estimated by the cavity method. 
The scaling parameters $x$ and $y$ are estimated using a kernel method based on Gaussian process regression~\cite{harada2015kernel} under fixed transition temperature. 
Data with sizes ranging from $N=64$ to $1024$ are well-fitted to the scaling plot near the transition temperature even when $\gamma$ and $c$ are changed, indicating that the transition temperature of the cavity method is consistent with the MCMC results. 
The estimated exponents for the correlation size are $y=2.19(4)$ for $\gamma=\infty$ and $2.20(5)$ for $\gamma=0.505$ at $c=15$. 
The values of $y$ are similar for different $\gamma$ and $c$, but the value of $x$ varies considerably. 
It is naively expected that the MVC on the sparse graph belongs to the mean-field universality class, but we cannot conclude this definitely from the present results. 
The universality class, including dynamic behavior, will be discussed elsewhere. 

\subsection{Overlap distribution}
Next, we examine the distribution of overlap $P(q)$ in finite-size systems to discuss signs of RSB.
Suppose we consider the low-temperature limit as an extreme case for MVC. 
In the region of RS, the ground state consists of at most one backbone and many nonbackbone vertices.
The overlap distribution $P(q)$ has a single peak near the density of the backbone and is broadened by the nonbackbone contribution. 
In the RSB region of $c>e$, it has been shown experimentally that the ground states can be decomposed into multiple clusters separated by a Hamming distance greater than 2~\cite{barthel2004clustering}.
If we define ``backbone'' and ``nonbackbone'' for each decomposed cluster, $P(q)$ will have multiple peaks caused by the overlap between different ``backbones'' and peaks caused by overlap within the same ``backbone''.

\begin{figure}[t]
    \includegraphics[width=0.95\linewidth]{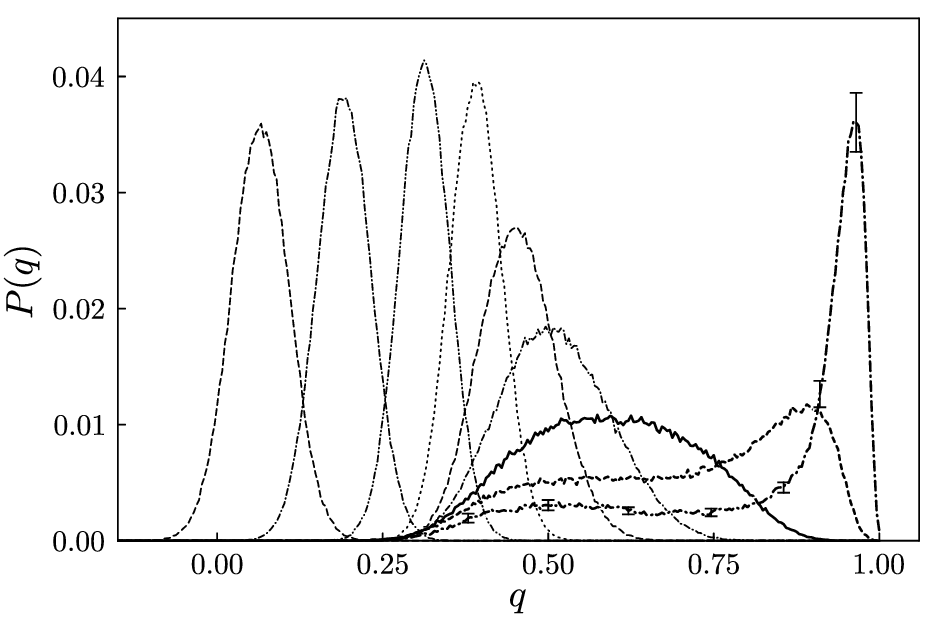}
    \caption{\label{fig:pofq_c5g11}Graph-averaged overlap distribution $P(q)$ at several $\beta$s for $N=512$ with mean degree $c=15$ and $\gamma=1.1$. $P(q)$ is ordered by $\beta$ from left to right, $\beta=0.1, 0.28, 0.62, 1.29, 2.37$, and $3.06$ below $\beta_c=3.89$, and $3.98, 5.64$, and $10.0$ above $\beta_{\mathrm{c}}$. The results are averaged over 200 random graphs, and the error bars evaluated by the bootstrap method are plotted only at $\beta=10$ for visibility.
    }
\end{figure}

\begin{figure}[t]
    \includegraphics[width=0.95\linewidth]{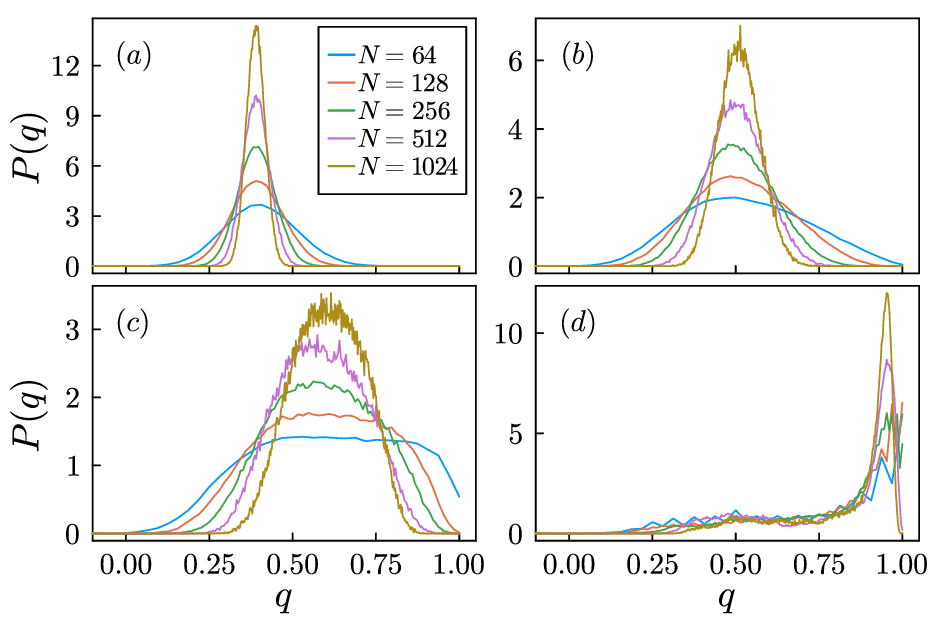}
    \caption{\label{fig:pofq_size}(Color online) System-size $N$ dependence of the graph-averaged overlap distributions $P(q)$ for mean degree $c=15$ and $\gamma=1.1$.
    Temperatures are at (a) $\beta=1.05$, (b) 3.06, (c) 3.98, and (d) 10.0, and (c) and (d) are above the inverse transition temperature $\beta_c=3.89$. The average of the random graphs is taken at approximately $200$. 
    }
\end{figure}

Here, the behavior of $P(q)$ is discussed, focusing on the case $c=15$ and $\gamma=1.1$, where the critical temperature $\beta_c=3.89$.
The $\beta$ dependence of $P(q)$ for $N=512$ is shown in Fig.~\ref{fig:pofq_c5g11}. 
It can be seen that the distribution follows a Gaussian distribution at a sufficiently high temperature, and below the transition temperature, the distribution has a large peak and a tail with a small overlap. 
The $N$ dependence of $P(q)$ at four different temperatures above and below $\beta_{\mathrm{c}}$ is also shown in Fig.~\ref{fig:pofq_size}.
As can be seen, below $\beta_{\mathrm{c}}$, $P(q)$ approaches a Gaussian distribution as $N$ increases, while above $\beta_{\mathrm{c}}$, it converges to each distribution significantly different from a Gaussian distribution.
This is considered to be a sign of RSB at finite temperature.
Figure~\ref{fig:pofq_size}(d) shows that $P(q)$ at $\beta=10$, well below the transition temperature, has a long tail that does not tend to disappear even when $N$ is increased.
This temperature is sufficiently low for the probability of occurrence of the ground state to be large.
The strong peak just below $q=1.0$ in $P(q)$ is considered to be an overlap between states belonging to the same cluster as discussed above, and the long tail starting from $q\approx 0.25$ is a result of an overlap between states belonging to different clusters.

\begin{figure}[t]
    \includegraphics[width=0.9\linewidth]{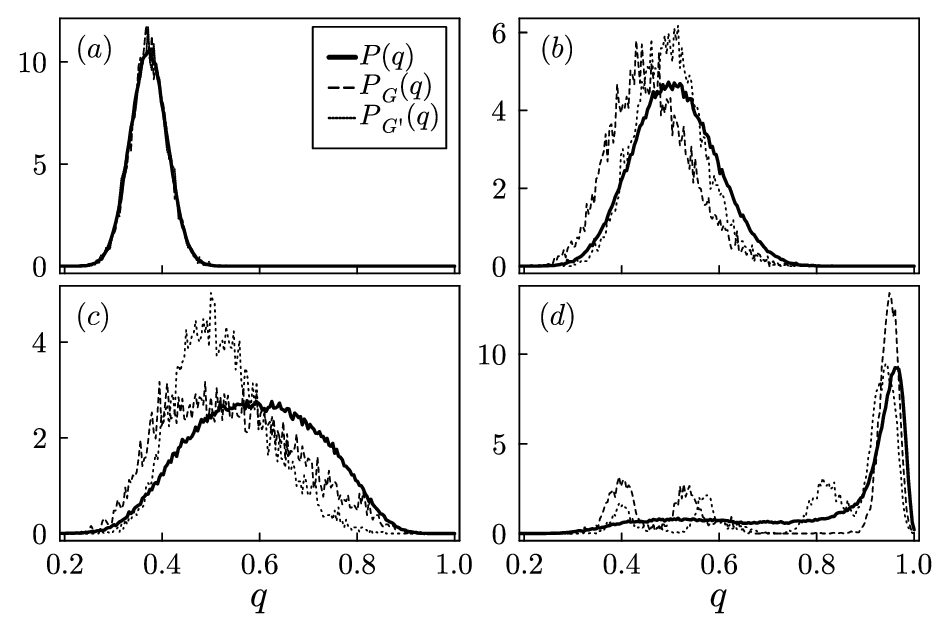}
    \caption{\label{fig:pofq_self_ave}Graph averaged overlap distribution $P(q)$ (solid line) and $P_G(q)$ for two sample graph instances (dashed and dotted lines) for $N=512$ with mean degree $c=15$ and $\gamma=1.1$.
    Temperatures are at (a) $\beta=1.05$, (b) 3.06, (c) 3.98, and (d) 10.0. Temperatures in (c) and (d) are above the inverse transition temperature $\beta_c=3.89$. 
    }
\end{figure}

One of the interesting properties resulting from RSB in the mean-field theory of spin glasses is the lack of self-averaging. 
In the context of optimization problems, this means that fluctuations caused by instances remain significantly in the thermodynamic limit. 
To examine the self-averaging property, the distribution $P_G(q)$ of two different randomly chosen graphs $G$ and $G'$ is plotted together with the graph-averaged $P(q)$ in Fig.~\ref{fig:pofq_self_ave}.
At a sufficiently high temperature in Fig.~\ref{fig:pofq_self_ave}(a) $\beta=1.05$, the two $P_G(q)$ almost coincide with $P(q)$, indicating that the self-averaging property is satisfied. 
However, at lower temperatures, as shown in Figs.~\ref{fig:pofq_self_ave}(c) and (d), their $P_G(q)$ have different distributions, suggesting a lack of self-averaging.
Particularly, for $\beta=10.0$, besides a strong peak near $q=1$, each $P_G(q)$ has two or three isolated small peaks at different positions.
This result is consistent with the previous study~\cite{barthel2004clustering}, which shows that the optimal solutions of MVC consist of a small number of clusters.
This suggests that the structures of the clusters differ significantly for each individual graph and that the long tail in $P(q)$ is the superposition of the overlaps of these small isolated peaks.
Even in Fig.~\ref{fig:pofq_self_ave}(b), which is slightly higher than the transition temperature, $P_G(q)$ does not coincide with $P(q)$, but this is due to the finite-size effect, and we expect self-averaging to hold for $N\to\infty$.

\section{\label{sec:conclusion}Conclusion}

In this study, we formulate MVC with the penalty method to account for both feasible and infeasible solutions. 
We conduct a statistical-mechanical analysis of typical properties of MVC on ER random graphs with the aim of investigating the effect of constraint relaxation on constrained combinatorial optimization problems.
Using the replica method under the assumption of replica symmetry, the condition for obtaining feasible solutions in the low-temperature limit is naturally derived, and the complex structure of the ground states in infeasible conditions is revealed. 
This structure is determined by the balance between the cost and penalty terms to the energy function and requires a careful treatment of the undetermined constraints as well as the nonbackbone vertices discussed in the previous study.
As a byproduct of this analysis, it was found that correctly incorporating their effects further improves the accuracy of the RS approximation of the minimum cover ratio in the RSB region, which no longer breaks the combinatorial lower bound known previously. 
Although not addressed in this paper, there are several efforts to improve the accuracy of the approximation using the 1-RSB solution~\cite{zhang2009stability}, and their relationship to our RS analysis will need to be investigated in the future.

In the low-temperature limit, the effective field distributions exhibit a discrete nature, such as the integer and rational ansatz, which can be understood as the result of competing cost and penalty functions in the ground state. 
This is intuitively due to the discrete number of adjacent variables. 
Such an interpretation may be universally applicable to the ground states of other combinatorial optimization problems. 
For example, the integer ansatz, often studied in MVC and other literature, is typical for problems with unit penalty coefficients. 
When dealing with arbitrary coefficients, the effective field can take nonzero values only in the linear combination of these coefficients.
In the constraint violation region, the penalty function displays a staircase-like structure and, importantly, behaves as a nonincreasing function with respect to the penalty coefficients. 
This feature is expected to be prevalent in constrained combinatorial optimization problems.
Understanding this structure might lead to the development of more sophisticated adaptive penalty function methods.

The RS/RSB phase boundary in the penalty method of MVC was also obtained from the stability analysis of the RS solution and the divergence of the spin-glass susceptibility $\chi_{\rm SG}$ using the cavity method. 
We find that the constraint relaxation leads to a decrease of the critical temperature $T_{\mathrm{c}}$ and an extension of the stability limit of the RS solution in the low-temperature limit.
These analytical results are also justified by calculations of $\chi_{\rm SG}$ via the MCMC method and its finite-size scaling analysis.
Combined with the results from the MCMC method, the individual properties of each instance do not appear to be pronounced at higher temperatures, where the RS solution becomes stable. 
In other words, the equilibrium states are expected to be easy to reach, and the fact that the constraint relaxation shifts its phase boundary to lower temperatures means that such an easy-to-reach region is extended. 
In this sense, the constraint relaxation is effective in solving optimization problems. 
By contrast, the penalty region where the transition temperature is remarkably lowered is also the region where infeasible solutions are more pronounced, and the trade-off relationship can be understood in terms of this phase diagram. 

Analyzing the effect of constraint relaxation on the solution time of combinatorial optimization problems is an interesting ultimate goal of our research.
The penalty strength $\gamma$ is a parameter that can be tuned when solving the problem, and we hope that our results will help to improve methods such as the exchange Monte Carlo and adaptive penalty methods, as well as to develop new algorithms.

\begin{acknowledgments}
This work was supported by JST Grant No. JPMJPF2221 and JSPS KAKENHI Grant No. 23H01095.
\end{acknowledgments}

\appendix

\section{Replica symmetric calculation of MVC with penalty term}

Here, we present the derivation of the self-consistent equation in Eq.~(\ref{eq:self-consistent}) for the effective-field distribution and the formulas of the cover and penalty ratios under replica symmetric ansatz.

\begin{widetext}
\subsection{Derivation of self-consistent equation for $p(h)$ of Eq.~(\ref{eq:self-consistent})}\label{sec:self_consistent_eq}
The formal free-energy density $g(\{C(\vec{\xi})\})$ with the order parameter $C(\vec{\xi})$ is given by Eq.~(\ref{eq:freeentropyfunction}), and the purpose of this subsection is to derive the self-consistent equation for the distribution of the effective field $p(h)$, Eq.~(\ref{eq:self-consistent}), under the RS ansatz. The self-consistent equation of $C(\vec{\xi})$ is obtained by differentiating Eq.~(\ref{eq:freeentropyfunction}) as 
\begin{equation}
C(\vec{\xi}) = \exp\left(-1+\zeta-\beta\mu \vec{1}\cdot\vec{\xi}
-c\sum\nolimits_{\vec{\xi'}}C(\vec{\xi'})\left(1-\exp\left(-\beta\gamma(\vec{1}-\vec{\xi})\cdot(\vec{1}-\vec{\xi'})\right)\right) \right), \quad \forall\vec{\xi}, 
\label{eqn:saddle-point}
\end{equation}
where $\zeta$ is the Lagrange multiplier to impose the normalization condition, $\sum_{\vec{\xi}} C(\vec{\xi})=1$. 
The formulation is approximately identical to that in the previous study~\cite{weigt2001minimal}, and the only difference is the last term in the exponential term of Eq.~(\ref{eqn:saddle-point}) due to the penalty function. 
Using the order parameter $C_{\rm RS}$ under the RS ansatz given by Eq.~(\ref{eqn:eff_potential}), this exponential term can be calculated as
\begin{align*}
\sum_{\vec{\xi'}}C_{\rm{RS}}\left(\sum_\alpha\xi'^{(\alpha)}\right)\left(1-e^{-\beta\gamma(\vec{1}-\vec{\xi})\cdot(\vec{1}-\vec{\xi'})}\right)
    &= 1-\int\frac{dh p(h)}{(1+e^{-\beta\mu h})^n}\sum_{\vec{\xi'}}\prod_\alpha e^{-\mu h \xi'^{(\alpha)}-\beta\gamma(1-\xi^{(\alpha)})(1-\xi'^{(\alpha)})}\\
    &= 1-\int dhp(h)\left(\frac{1+e^{-\beta\mu h}}{e^{-\beta\gamma}+e^{-\beta\mu h}}\right)^{\sum_\alpha \xi^{(\alpha)} - n}.
\end{align*}
Substituting Eq.~(\ref{eqn:eff_potential}) into both sides of Eq.~(\ref{eqn:saddle-point}), we obtain 
\begin{align}
    \int dh p(h)\frac{\exp(-\beta\mu h\sum_\alpha \xi^\alpha)}{(1+e^{-\beta\mu h})^n}
     & 
    = \exp\left(-c-1+\zeta-\beta\mu\sum_{\alpha}\xi^{(\alpha)}+c
    \int dh p(h) \left(\frac{1+e^{-\beta\mu h}}{e^{-\beta\gamma}+e^{-\beta\mu h}}\right)^{\sum_\alpha \xi^{(\alpha)} - n}
    \right). 
\end{align}
Furthermore, putting $y=\sum_{\alpha}\xi^{(\alpha)}$ and taking the replica limit of $n\to 0$, we obtain
\begin{align*}
    \int dh p(h) e^{-\beta\mu hy}
     & 
    = e^{-c-\beta\mu y}
    \sum_{l=0}^\infty \frac{c^l}{l!}\int p(h_1)dh_{1}\cdots p(h_l)dh_l
    \exp\left(y\sum_{i=1}^l\log\left(\frac{1+e^{-\beta\mu h_i}}{e^{-\beta\gamma}+e^{-\beta\mu h_i}}\right)
    \right),
\end{align*}
\end{widetext}
where the Lagrange multiplier is eliminated by explicitly imposing the normalization condition. 
When this equation is regarded as a Laplace transform, the self-consistent equation of $p(h)$ is derived as Eq.~(\ref{eq:self-consistent}) by the inverse Laplace transform.

\subsection{Derivation of Eqs.~(\ref{eq:coverratio}) and (\ref{eq:penaltyratio}) for RS expression of the cover and penalty ratios}\label{sec:cover_and_penalty_ratio}
As shown in Eqs.~(\ref{eq:cost_density}) and (\ref{eq:penalty_density}), the cover ratio $\rho$ and the penalty ratio $\nu$ are derived from the free entropy function of Eq.~(\ref{eq:freeentropyfunction}), which are given by
\begin{align}
    \rho(\beta,\gamma) 
    &=\lim_{n\to0}\frac{1}{n}\sum\nolimits_{\vec{\xi}}\hat{C}(\vec{\xi})\vec{1}\cdot\vec{\xi},\label{eq:coverratioA} \\
    \nu(\beta,\gamma)
    &=\lim_{n\to0}\frac{1}{n}\frac{c}{2}\sum\nolimits_{\vec{\xi},\vec{\xi}'} \hat{C}(\vec{\xi}) \hat{C}(\vec{\xi}')\nonumber \\
    &\times (\vec{1}-\vec{\xi})\cdot(\vec{1}-\vec{\xi}')e^{-\beta\gamma(\vec{1}-\vec{\xi})\cdot(\vec{1}-\vec{\xi}')},
    \label{eqn:penaltyA}
\end{align}
respectively. 
Substituting Eq.~(\ref{eqn:eff_potential}) into Eq.~(\ref{eq:coverratioA}), the RS cover ratio $\rho$ is obtained as Eq.~(\ref{eq:coverratio}) 

Under the RS ansatz, the right-hand side of Eq.~(\ref{eqn:penaltyA}) can also be calculated explicitly. The right-hand side before taking the replica limit, denoted by $\nu_n$, is expressed as 
\begin{align*}
 \nu_n &=\frac{c}{2}\sum_{\vec{\xi},\vec{\xi}'}\int dhdh'p(h)p(h') \frac{e^{-\beta h\sum_{\alpha} \xi^{(\alpha)}}e^{-\beta h'\sum_{\alpha} \xi'^{(\alpha)}}}{(1+e^{-\beta h})^n(1+e^{-\beta h'})^n} \\
 &\times \sum_{\alpha'}(1-\xi^{(\alpha')})(1-\xi'^{(\alpha')}) \prod_{\alpha} e^{-\beta\gamma(1-\xi^{(\alpha)})(1-\xi'^{(\alpha)})} \nonumber \\
  &=\frac{c}{2}\int dhdh' \frac{p(h)p(h')}{(1+e^{-\beta h})^n(1+e^{-\beta h'})^n} \\
&\times ne^{-\beta\gamma} (e^{-\beta\gamma}+e^{-\beta h}+e^{-\beta h'}+e^{-\beta(h+h')})^{n-1},
\end{align*}
where the sum of $\vec{\xi}$ and $\vec{\xi}'$ is calculated.
Finally, taking a replica limit, the RS penalty ratio of Eq.~(\ref{eq:penaltyratio}) is derived.

\section{Population dynamics method and its time evolution at zero temperature}
In this appendix, we describe the population dynamics method for numerically solving the self-consistent or recursive equations and then discuss its solution in the low-temperature limit.

\subsection{Population dynamics method}\label{sec:population_dynamics}

The population dynamics method~\cite{mezard2001bethe} is an iterative solution method to find the solution of Eq.~(\ref{eq:self-consistent}) by approximating the distribution $p(h)$ by the sample set $\{h_i\}$.
The procedure is shown in Algorithm~\ref{alg:popdyn1}.
\begin{algorithm}[H]
    \caption{Population dynamics for $p(h)$} \label{alg:popdyn1}
    \begin{algorithmic}
        \Require $c,\beta,\gamma; N_{\rm pop},N_{\rm itr}$
        \State \textbf{Initialization:} $ P_{\rm prev} = \{h_i\}_{i=1}^{N_{\rm pop}}, P_{\rm new}\gets P_{\rm prev}$
        \For{$itr=1$ to $N_{\rm itr}$}
            \For{$i=1$ to $N_{\rm pop}$ }
                \State $l\sim \text{Poisson}(c)$
                \State draw $l$ indices $\{i_j\}_{j=1}^l \in \{1\dots N_{\rm pop}\}$
                \State update $h_i'$ in $P_{\rm new}$ as: $h_i' \gets 1 - \sum_{j=1}^l K(h_{i_j};\beta,\gamma)$
            \EndFor
            \State $P_{\rm prev} \gets P_{\rm new}$
        \EndFor \\
        \Return $P_{\rm new}$
    \end{algorithmic}
\end{algorithm}

An arbitrary sample set $\{h_i\}$ may be used as the initial distribution for $P_{\rm prev}$ and $P_{\rm new}$.
At each iteration in the inner for-loop, $l$ is drawn from the Poisson distribution with mean $c$, $P(l)=e^{-c}c^l/l!$, then, $l$ samples are drawn from $P_{\rm prev}$ randomly. 
A new sample $h_i'$ is calculated by the equation in the delta function in Eq.~(\ref{eq:self-consistent}).
Repeating this procedure for the number of samples $N_{\rm pop}$ generates a new distribution $P_{\rm new}$, which then replaces an old distribution $P_{\rm prev}$.

The above iteration ensures that $\{h_i\}$ converges to the solution $p(h)$ in the limit $N_{\rm pop}\to\infty$~\cite{mezard2001bethe}.
In this pseudo-code, the number of iterations is given by $N_{\rm itr}$, and its convergence can be confirmed using a metric such as Kolmogorov-Smirnov distance~\cite{an1933sulla} between $P_{\rm prev}$ and $P_{\rm new}$.

As discussed in Sec.~\ref{sec:damping}, the distribution of ${h_i}$ can oscillate depending on the values of $c$, $\beta$, and $\gamma$. 
To mitigate this, a method known as damping is employed, where only a fraction of $\{h_i\}$, denoted as $0 < r < 1$, is updated. 
Specifically, when generating the next $P_{\rm new}$ in the inner for-loop, only $rN_{\rm pop}$ samples are updated.

When applying the population dynamics method to a distribution such as $p(h,\tilde{h})$ in Eq.~(\ref{eq:self-consistent-pofhh}) or $p(h,\Delta)$ in Sec.~\ref{sec:div_of_chiSG}, it is important to update two variables simultaneously.
Specifically, in the case of $p(h,\tilde{h})$, new $h_i'$ and $\tilde{h}_i'$ are generated using $l$ identical samples $h_j$s drawn from $P_{\rm prev}$, as shown in Algorithm~\ref{alg:popdyn2}.
\begin{algorithm}[H] 
    \caption{Population dynamics for $p(h,\tilde{h})$} \label{alg:popdyn2}
    \begin{algorithmic}
        \Require $c,\beta,\gamma; N_{\rm pop},N_{\rm itr}$
        \State \textbf{Initialization:} $ P_{\rm prev} = \{(h_i,\tilde{h}_i)\}_{i=1}^{N_{\rm pop}}, P_{\rm new}\gets P_{\rm prev}$
        \For{$itr=1$ to $N_{\rm itr}$}
            \For{$i \in 1:N_{\rm pop}$ }
                \State $l\sim \text{Poisson}(c)$
                \State draw $l$ indices $\{i_j\}_{j=1}^l \in \{1\dots N_{\rm pop}\}$
                \State update $h_i'$ in $P_{\rm new}$ as: $h_i' \gets 1 - \sum_{j=1}^l K(h_{i_j})$
                \State update $\tilde{h}_i'$ in $P_{\rm new}$ as: $\tilde{h}_i' \gets \sum_{j=1}^l \tilde{K}(\tilde{h}_{i_j}|h_{i_j})$
            \EndFor
            \State $P_{\rm prev} \gets P_{\rm new}$
        \EndFor \\
        \Return $P_{\rm new}$
    \end{algorithmic}
\end{algorithm}

\subsection{Derivation of integer ansatz, irrational and rational ans\"{a}tze at zero-temperature} \label{sec:PD_at_zero_temp}

Here we show the validity of ans\"{a}tze for $p(h)$ in low-temperature limit in terms of the time evolution of the population dynamics method. 
First, the integer ansatz for $\gamma \ge 1$ is discussed in detail.
Then, the irrational and rational ansatz for $0<\gamma<1$ can be explained in the same manner as for the integer ansatz. 

A new sample is generated as $h_i' = 1 - \sum_{j=1}^l K(h;\infty,\gamma)$ which represents a mapping from $l$ samples $\{h_j\}$ to a sample $h_i'$.
The function $K(h;\infty,\gamma)$ reads,
\begin{equation}
  K(h;\infty,\gamma) = \begin{cases}
       \gamma & (\gamma \le h) ,\\
       h      & (0 < h < \gamma) ,\\
       0      & (h\le0) .
      \end{cases} 
      \label{eq:Kofh_inftyA}
\end{equation}
For $\gamma>0$, any $h_j<0$ does not contributes to new sample $h_i'$.

\begin{figure}[t]
    \centering
    \includegraphics[keepaspectratio, width=0.6\linewidth]{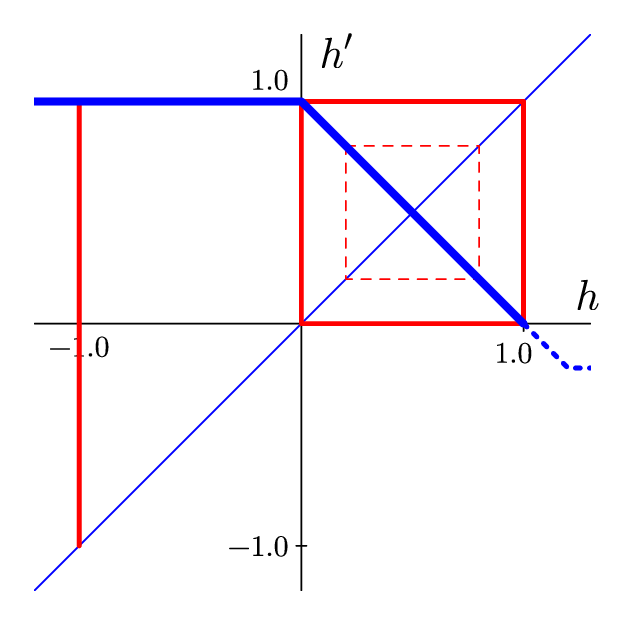}
    \caption{\label{fig:dyn_popdyn}(Color online) Dynamical mappings for the population dynamics at the low-temperature limit in the feasible region. The (blue) thick line represents the mapping, $h'=1-K(h)$, for $l=1$, and the (red) horizontal and vertical lines represent the paths that $h_i'$ is generated from $\{h_j\}$. The cycles shown by the dotted lines eventually disappear because their population is not supplied by other $h_j$, and then values other than $h_j=0$ and $1$ disappear in the range $0\le h \le1$.
    }
\end{figure}

We examine the integer ansatz in the case of $\gamma \ge 1$. 
First, when $l=0$ or the sample $\{h_j\}$ are all less than zero, $h_i'=1$ is generated.
Second, when only one of $l$ samples $\{h_j\}$ with $l\geq 1$ is larger than zero, $h_i' \gets 1-h_j$.
The mapping function for $l=1$ is shown in Fig.~\ref{fig:dyn_popdyn}.
Furthermore, considering the case when two or more $h_j$'s are larger than zero, $h_i'=1-\sum_j h_j$. 
Therefore, any $h_i'>0$ is given by $1-\sum_jh_j$ with $0\le h_j\le 1$.
The above process can be summarized as consisting of three flows: (a) a looping flow within the range of $0 \le h_j \le 1$, (b) an outflow process from the sum of $0 \le h_j \le 1$ to $h_i'<0$, and (c) a process of generating $h_i'=1$ from multiple $h_j<0$ or nothing ($l=0)$. 
Since samples in the range $0 < h_j < 1$ are derived only from the prepared initial distribution, this population decreases by the outflow process described in (b), and eventually disappears after a sufficient number of iterations. 
As a result, in the population dynamics of $\beta\to\infty$, only 0 and 1 remain in the range of $0\le h_j \le1$, and therefore, $\{h_i'\}$ can take only integer values less than 1.

Note that when the damping in the population update is performed, there may remain real-valued samples $h_i$ even after long iterations, but they should always disappear as no new population is supplied, and it is recommended to use only integers or 0 for the initial value for $\beta\to\infty$.

\begin{figure}
    \centering
    \includegraphics[keepaspectratio, width=0.6\linewidth]{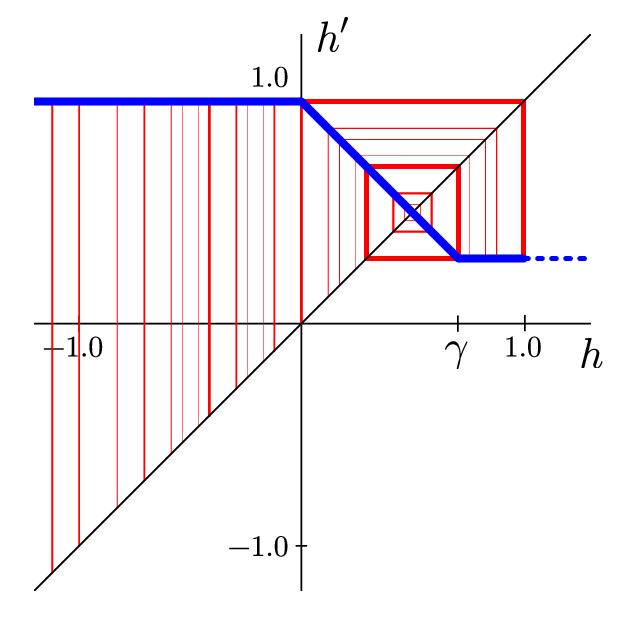}
    \caption{\label{fig:popdyn_infeasble}(Color online) Dynamical mappings for the population dynamics at the low-temperature limit in the infeasible region. 
    There exists a stable cycle of $\gamma\leftrightarrow 1-\gamma$ whose population is provided from other $h_j$'s.
    Then, there are innumerable secondary $h_i'$ derived from their sums.
    Any value other than $h=1-l-l'\gamma,\quad l,l'\in\mathcal{Z}$ eventually vanishes for the same reason as the dotted cycles in Fig.\ref{fig:dyn_popdyn}.}
\end{figure}

The irrational ansatz in Eq~(\ref{eq:pofh_zero_T_solution_irrational}) for $0<\gamma<1$ is slightly complicated.
The mapping function shown in Fig.~\ref{fig:popdyn_infeasble} allows many samples to be gathered into the loop-like flow consisting of $ h_j\in\{1-\gamma, \gamma, 1\}$. 
Since only the values generated from the above three values survive, the only possible values for $h_i'$ are the sum of an integer and an integer multiple of $\gamma$, that is, $\{h_i\mid h_i=1-l-\gamma l' \le1, l,l'\in \mathbb{Z}\}$.
When $\gamma$ is a rational number, it can also be observed that the rational ansatz in Eq.~(\ref{eq:pofh_zero_T_solution_rational}) holds, because the sum of an integer and a rational number is a rational number.

\section{Derivation of the self-consistent equation with correlation field Eqs.~(\ref{eq:h_to_h_plus_h_tilde}) and (\ref{eq:self-consistent-pofhh})}\label{sec:calc_corr_field}

In this appendix, we provide a supplementary explanation of the treatment of the effective field distribution with a correction field. 
In Sec.~\ref{sec:correction_field}, we introduced fluctuations around a delta-function distribution in the effective-field distribution $p(h)$ at sufficiently low temperatures in Eq.~(\ref{pofh_evanescent}), and we assume that the width of these fluctuations scales with temperature and $p(h)$ at $\beta\gg1$ can be described by
\begin{equation}
    p(h,\tilde{h}/\beta) = \sum_{\{h'\}} r_{h'}\delta(h-h')\rho_{h'}(\tilde{h}/\beta) ,
    \label{eq:pofhtildeh_A}
\end{equation}
where the distribution $\rho_{h'}$ representing the fluctuations is normalized for any $h'$ as $\int d\tilde{h}/\beta \rho_{h'}(\tilde{h}/\beta) = 1$. 
The original $p(h)$ is reduced by marginalizing $p(h,\tilde{h}/\beta)$ to 
\begin{equation}
    p(h)=\int dh' \frac{d\tilde{h}}{\beta} \delta\left(h-h'-\frac{\tilde{h}}{\beta}\right) p(h',\tilde{h}/\beta) .
    \label{eq:h_to_h_plus_h_tilde_beta}
\end{equation}
This equation is equivalent to Eq.~(\ref{eq:h_to_h_plus_h_tilde}). 
Plugging this into Eq.~(\ref{eq:self-consistent}), we obtain 
\begin{align*}
    p(h,\tilde{h}/\beta) 
    =& e^{-c}\sum_{l=0}^\infty \frac{c^l}{l!} \int \prod_{i=1}^l dh_i \frac{d\tilde{h}_i}{\beta} p(h_i, \tilde{h}_i/\beta) \nonumber \\
    &\times 
  \delta\left(h+\frac{\tilde{h}}{\beta} -1 + \sum_{j=1}^l K(h_j+\frac{\tilde{h}_j}{\beta}) \right) .
\end{align*}
Expanding $K(h)$ assuming $\beta\gg1$ and leaving up to the $O(1/\beta)$ term, the delta-function term reads 
\begin{equation}
    \delta\left(h-1+\sum_{j=1}^l K(h;\infty,\gamma) + \frac{1}{\beta}\left(\tilde{h} + \sum_{j=1}^l \tilde{K}(\tilde{h} \mid h) \right)\right) .
    \label{eq:selfconsisient_eq_for_large_beta}
\end{equation}
where $K(h;\infty,\gamma)$ is the same as Eq.~(\ref{eq:Kofh_infty}).
The update function for $\tilde{h}$ depending on the positions $\{h\}$ of each peak is shown in Eq.~(\ref{eq:update_corr_field}).
For the above delta function to make a non-trivial contribution to any $\beta\gg1$, each $O(\beta^0)$ and $O(1/\beta)$ term must be zero. 
Therefore, using the normalization condition $\int (d\tilde{h}/\beta) \rho_{h'}(\tilde{h}/\beta) = 1$ and the property of the delta function, $\delta(\beta x)=\delta(x)/|\beta|$, we can divide the delta functions into a product and remove $\beta$'s from Eq.~(\ref{eq:selfconsisient_eq_for_large_beta}).  
Finally, the self-consistent equation for $p(h,\tilde{h})$ can be derived as Eq.~(\ref{eq:self-consistent-pofhh}), which can be regarded as a joint of the self-consistent equation with respect to $h$, Eq.~(\ref{eq:self-consistent}), and the self-consistent equation for $\tilde{h}$ conditioned on $h$. 
Since the $h$ part of Eq.~(\ref{eq:self-consistent}) is independent of $\tilde{h}$, the $h$-dependence of the solution is the same as that without the correction field $\tilde{h}$.
Therefore, it is justified to rescale Eqs.~(\ref{pofh_evanescent}) and (\ref{eq:h_to_h_plus_h_tilde}) as Eqs.~(\ref{eq:pofhtildeh_A}) and (\ref{eq:h_to_h_plus_h_tilde_beta}), respectively.

The cover and penalty ratios are straightforward to be calculated by just plugging Eq.~(\ref{eq:h_to_h_plus_h_tilde}) into Eq.~(\ref{eq:coverratio}) and (\ref{eq:penaltyratio}), respectively.
The low-temperature limit of the factor of the cover ratio is shown in Eq.~(\ref{eq:cov_factor_corrected}). 
The factor of the penalty ratio in Eq.~(\ref{eq:penaltyratio}) in the low-temperature limit is also evaluated as  
\begin{align}
  &\lim_{\beta\to\infty} \frac{1}{1+e^{-\beta (h-\gamma)-\tilde{h}}+e^{-\beta (h'-\gamma)-\tilde{h}'}+e^{-\beta(h+h'-\gamma)-\tilde{h}-\tilde{h}'}} \nonumber \\
  &= \begin{cases}
  1 & (h,h'>\gamma) ,\\
  1/(1+e^{-\tilde{h}}) & (h=\gamma,h'>\gamma \text{ or } h \leftrightarrow h') ,\\
  1/(1+e^{-\tilde{h}}+e^{-\tilde{h}'}) & (h=h'=\gamma) ,\\
  0 & (h<\gamma \text{ or } h'<\gamma) .
  \end{cases}
\end{align}
This yields the penalty ratio at $\gamma=1$ in Eq.~(\ref{eq:correct_penalty_ratio_at_g1}).


\bibliography{apssamp}

\end{document}